\documentclass[11pt]{article}
\usepackage[margin=1.0in]{geometry}
\usepackage{setspace}
\usepackage{amsmath,amssymb,graphicx}
\usepackage[pagebackref=true]{hyperref}
\usepackage{slashed}

\usepackage[T1]{fontenc}

\usepackage[utf8]{inputenc}

\usepackage[greek,english]{babel}

\usepackage{framed}

\usepackage[font=small,labelfont=bf]{caption}
\usepackage{cite}
\usepackage{comment}

\usepackage{slashed}

\usepackage{fancyhdr}

\fancypagestyle{firstpage}{
  \fancyhf{}
  \fancyhead[R]{MPP-2026-146}
  \fancyfoot[C]{\thepage}
  
}

\newcommand{\be}{\begin{equation}}
\newcommand{\ee}{\end{equation}}

\usepackage{xcolor}
\usepackage{bm}

\title{\bf  Taxonomy of Potentials in Asymptotic Limits}
\author{Muldrow Etheredge${}^1$, Dieter L\"ust${}^{1,2}$, Tom Rudelius${}^3$\\[0.25em]
{\small \color{gray} \texttt{muldrow@mpp.mpg.de},~\texttt{luest@mpp.mpg.de},}\\[0.25em] {\small \color{gray} \texttt{thomas.w.rudelius@durham.ac.uk}}\\[0.25em]
{\small${}^1$ Max-Planck-Institut f\"ur Physik (Werner-Heisenberg-Institut), Garching, 85748, GER}\\[0.25em]
{\small${}^2$ Arnold Sommerfeld Center for Theoretical Physics,}\\ 
{\small Ludwig-Maximilians-Universit\"at M\"unchen, M\"unchen, 80333, GER}\\[0.25em]
{\small${}^3$ Department of Mathematical Sciences, Durham University, Durham, DH1 3LE, UK}
}

\begin{document}
\maketitle

\thispagestyle{firstpage}

\begin{abstract}

In infinite-distance limits of scalar field moduli spaces in quantum gravity, particle masses, brane tensions, and scalar field potentials scale exponentially with geodesic distance. Previous work has shown that the exponential decay rates of particle masses and brane tensions admit a discrete classification \cite{Etheredge:2024tok, Etheredge:2025ahf}. In this work, we extend this classification to the case of scalar field potentials. We find that leading contributions to the potential typically scale with tensions of codimension-1 branes as $V \sim T_{d-1}^2$ or codimension-0 branes as $V \sim T_d$, and these relations hold formally even when the associated branes are absent from the spectrum. As a result, the taxonomy rules for potentials are intimately connected to the brane taxonomy rules. More generally, potential contributions can be labeled by a collection of integers, which correspond physically to their transformation properties under Weyl rescalings and their order in the string loop expansion. This implies that the vector $\vec v = - \vec \nabla \log V$ is lattice-valued, and indeed it lies in precisely the same lattice as the analogous vectors $\vec \alpha = - \vec \nabla \log T_d$ for codimension-0 branes of tension $T_d$. We verify our taxonomy rules in various examples in string theory. We show that these rules imply that sums of positive potential terms satisfy the Strong Asymptotic de Sitter Conjecture \cite{Obied:2018sgi, Hebecker:2018vxz, Rudelius:2021oaz, Rudelius:2021azq, Rudelius:2022gbz}, which requires $|\vec \nabla \log V| \geq 2/\sqrt{d-2}$ in asymptotic regimes of scalar field moduli space, and they imply that higher-derivative gravitational corrections are suppressed by powers of the species scale \cite{vandeHeisteeg:2022btw, vandeHeisteeg:2023dlw, Castellano:2023aum}.

\end{abstract}
\clearpage

\tableofcontents

\clearpage

\section{Introduction}

The landscape of quantum gravity remains shrouded in mystery, but a measure of clarity can be found in asymptotic regimes of scalar field moduli space. These regimes feature parametric control of perturbative expansions, which allows for rigorous exploration of their physics. Furthermore, they appear to exhibit universal features, which are encoded in an array of Swampland conjectures.

The most famous of these conjectures is the Distance Conjecture \cite{Ooguri:2006in}, which stipulates that infinite-distance limits in scalar field moduli spaces contain towers of light particles, whose masses decay exponentially with proper field distance, $m \sim \exp(-\alpha \phi)$. The Brane Distance Conjecture \cite{Etheredge:2024amg} extends this claim from particles to branes, requiring exponential decay of brane tensions.
The Asymptotic de Sitter Conjecture \cite{Obied:2018sgi, Hebecker:2018vxz} similarly holds that scalar field potentials must decay exponentially in infinite-distance limits. These conjectures admit various refinements and sharpenings \cite{Baume:2016psm, Klaewer:2016kiy, Bedroya:2019snp, Rudelius:2021oaz, Rudelius:2021azq, Lust:2019zwm, Etheredge:2022opl,Rudelius:2022gbz, Lanza:2021qsu, vandeHeisteeg:2022btw, Cribiori:2022nke,Cribiori:2023sch,  Calderon-Infante:2023ler, vandeHeisteeg:2023dlw, Castellano:2023jjt, vandeHeisteeg:2023uxj}, one of which is the celebrated Emergent String Conjecture \cite{Lee:2019wij},  which holds that every infinite-distance limit is either a decompactification limit or an emergent string limit in some duality frame. 

In \cite{Etheredge:2024tok}, the Emergent String Conjecture was used to initiate a systematic classification of infinite-distance limits. The central objects of study in this program are known as \emph{\textbf{$\alpha$-vectors}}, which may be defined for a particle of mass $m$ as
\be
\vec \alpha = - \vec \nabla \log m\,,
\label{alpha1}
\ee
where $\vec \nabla$ is the gradient with respect to moduli space.\footnote{Here and throughout this work, we work in $d$-dimensional Planck units, $8 \pi G_d = \kappa_d^2 = 1$. The metric used is the moduli space metric $g_{ij}$, defined in the low-energy Lagrangian density $\mathcal L=\frac 12 R-\frac 12g_{ij}(\partial \phi^i)\cdot (\partial \phi^j)+\dots $.} In infinite-distance limits, the norms and dot products of the $\alpha$-vectors of particle towers are constrained by the Emergent String Conjecture to satisfy certain relations, referred to as \emph{\textbf{taxonomy rules}}. In \cite{Etheredge:2025ahf}, these rules were generalized from particle towers to branes, whose $\alpha$-vectors are defined analogously to \eqref{alpha1} with the mass $m$ replaced by the tension $T$.

In this work, we generalize this taxonomy program still further to classify the exponential decay of the scalar field potential $V(\vec\phi)$ in asymptotic regimes of scalar field space. In general, this potential consists of a sum of terms, $V(\vec\phi) = \sum_a V_a(\vec \phi)$, each of which decays exponentially with a linear combination of the fields $\vec \phi$:
\be
V_a \sim \exp(- \vec c_{a} \cdot \vec \phi)\,.
\ee
From this, we may define the $v$-vector of a given term in the potential as\footnote{Note that this definition works also for negative potential terms, $V_a < 0 $, since $- \vec \nabla \log(-V_a) = - \vec \nabla (i \pi + \log V_a) =- \frac{\vec \nabla V_a}{V_a}$, which does not depend on the sign of $V_a$.}
\be
\vec v_{a} = - \vec \nabla \log V_{a} = \vec c_a\,.
\ee
In this paper, we will show that these $v$-vectors obey taxonomy rules similar to those of the $\alpha$-vectors for particles and branes in \cite{Etheredge:2024tok, Etheredge:2025ahf}.

More precisely, we will show in \S\ref{sec:LEADING} that the $v$-vectors of leading contributions to the potential are typically associated with either codimension-1 branes or codimension-0 branes, with respective scaling relations
\be
V \sim T_{d-1}^2~~~~~\text{or}~~~~~V \sim T_{d}\,.
\ee
As a result, the $v$-vector is related to the $\alpha$-vectors of the respective codimension-1 or codimension-0 branes via
\be
\vec v = 2 \vec \alpha_{d-1}
~~~~~\text{or}~~~~~\vec v = \vec \alpha_{d}\,.
\label{vvsalphas}
\ee
We will argue that the relations \eqref{vvsalphas} hold at a formal, mathematical level even when the associated branes are absent from the physical spectrum. 
These relations allow us to infer the taxonomy rules of $v$-vectors from the taxonomy rules of $\alpha$-vectors for codimension-1 and codimension-0 branes. 

In \S\ref{sec:LEADING}, we will argue more generally that the $v$-vector of a term in the asymptotic potential can be expressed uniquely as an affine linear combination of a vector of non-negative integers $\vec w$, which encode the transformation properties of the potential term under Weyl transformations in diverse dimensions, and an integer $k_g \geq -2$ that encodes the order of the potential term in the string loop expansion. This implies that any $v$-vector necessarily resides in a lattice. We will further argue that this lattice may be identified precisely with the lattice generated by $\alpha$-vectors of codimension-0 branes $\{ \vec\alpha_d\}$, which was characterized in \cite{Etheredge:2025ahf} and notably contains all vectors of the form $\vec v = 2 \vec \alpha_{d-1}$, as required for consistency with \eqref{vvsalphas}. The upshot of this may be succinctly summarized as follows: given a canonically normalized radion $\rho$ associated with decompactifying a $d$-dimensional theory to a $D$-dimensional theory, the asymptotic scaling of the potential is of the form
\begin{align}
    V\sim \exp\left(-\frac{d(D-2)-P_\rho(d-2)}{\sqrt{(D-d)(D-2)(d-2)}}\rho\right),
\end{align}
where $P_\rho$ is an integer. For a canonically normalized $d$-dimensional dilaton $\phi$, the asymptotic exponential scaling is 
\begin{align}
    V\sim \exp\left(\left[\frac{d}{\sqrt{d-2}}+\frac{\sqrt{d-2}}2(d-P_\phi)\right]\phi\right),
\end{align}
where $P_\phi$ is an integer. We will further argue that these integers are bounded above as $P_{\rho} \leq D+2$, $P_{\phi} \leq d+2$. 

In \S\ref{sec:Relation}, we use our taxonomy rules to derive several well-studied Swampland bounds. First, we establish the Strong Asymptotic de Sitter Conjecture \cite{Rudelius:2021oaz,Rudelius:2021azq, Rudelius:2022gbz}, which places a sharp lower bound on the exponential decay rate of the potential in asymptotic regimes, $| \vec \nabla \log V| = || \vec v|| \geq 2/\sqrt{d-2}$. Notably, this result applies to multi-field potentials comprised of sums of positive terms. Next, we compare the decay rates of various mass scales in FRW cosmologies governed by fields rolling in asymptotic potentials. We demonstrate that for sums of positive potential terms obeying our taxonomy rules, the Hubble scale remains below the species scale $\Lambda_{\rm QG}$ \cite{Veneziano:2001ah, Dvali:2007hz, Dvali:2009ks, Dvali:2010vm}, as required for consistency of the effective field theory \cite{Hebecker:2018vxz}. 
We further show that our taxonomy rules imply that Wilson coefficients $c_k$ of higher-derivative gravitational corrections $c_k \mathcal{R}^k \subset \mathcal{L}$ scale with the species scale as 
\be
c_k \sim \Lambda_{\rm QG}^{2-2k}\,,
\ee
as previously conjectured in \cite{vandeHeisteeg:2022btw, vandeHeisteeg:2023dlw, Castellano:2023aum}.

In \S\ref{sec:CONC}, we conclude with a summary of our work and a discussion of directions for future study.

\section{Review of Brane Taxonomy}\label{sec:REVIEW}

In this section, we briefly review the taxonomy program developed in \cite{Etheredge:2024tok,Etheredge:2025ahf}.

The basic idea of the taxonomy program is to classify the possible exponential decay rates of particle and brane tensions in infinite-distance limits. 
To this end, let us consider a geodesic $\gamma$ traveling in an asymptotic region of moduli space in a theory in $d=D_1$ spacetime dimensions. According to the Emergent String Conjecture \cite{Lee:2019wij}, the geodesic will either approach a perturbative string limit or decompactify to $D_2$-dimensions. In either case, there will be a tower of particles with mass scale $m_1$; in the case of a perturbative string limit, these will be string oscillation modes, while in a decompactification limit, they will be Kaluza-Klein (KK) modes.

Suppose first that the limit in question is a decompactification limit. Then, in the asymptotic limit, the geodesic either terminates in the moduli space of the $D_2$-dimensional theory or it continues, yielding an infinite-distance limit in the moduli space of a theory in $D_2$ dimensions.
Applying the Emergent String Conjecture once again, we deduce that this limit must represent either a perturbative string limit in the $D_2$-dimensional theory, or else it must decompactify yet again to a $D_3$-dimensional theory. In these cases, there is respectively either a tower of string oscillator modes or a tower of KK-modes with mass scale $m_2>m_1$ (in $d$-dimensional Planck units).

Iterating this process, we generically find a sequence of decompactification limits of increasing dimensionality:
\be
d \equiv D_1 \rightarrow D_2 \rightarrow D_3 \rightarrow ... \rightarrow D_N \equiv D\,.
\ee
This process ends when either the geodesic terminates in a moduli space with no infinite-distance limits (such as in M-theory, which has no moduli space) or else the geodesic reaches a perturbative string limit.

Such a geodesic produces a hierarchy of mass scales from the KK-modes and possible string oscillator modes, $m_1<m_2<\dots<m_N$ (in $d$-dimensional Planck units). Such towers are called \emph{\textbf{principal towers}} \cite{Etheredge:2024tok}. These towers can also be viewed as exponentiations of radions and dilatons, and the logarithms of these masses span an asymptotically flat slice of moduli called the \emph{\textbf{principal plane}} \cite{Etheredge:2024tok}.

However, as explained in \cite{Etheredge:2024tok}, the logarithms of the masses of the principal towers $\log m_i$ do not form orthonormal coordinates on the principal plane. Defining the $\alpha$-vector of a tower as $(-\vec \nabla \log m)$, the non-orthonormality of these coordinates is captured by the \emph{\textbf{principal tower product rules}}. In particular, the $\alpha$-vectors $\vec \alpha_i$ and $\vec \alpha_j$ for two principal towers generically satisfy
\begin{align}
    \vec \alpha_i\cdot \vec \alpha_j=\frac 1{d-2}+\frac 1{n_i}\delta_{ij},
    \label{principaltower}
\end{align}
where in the case of a KK tower, $n_i = D_{i+1}-D_i$ is the dimension of the $i$th compactification manifold. If the principal tower is a tower of string oscillator modes, we formally set $n_i=\infty$.

These principal tower product rules can be violated in the presence of strong warping, which can cause the principal tower vectors $\vec \alpha_i$ to slide as a function of position in moduli space \cite{Etheredge:2023odp, Raucci:2026fzp}. However, \cite{Etheredge:2024tok} argued that violations of these product rules due to sliding occur only at a measure zero subset of the full space of infinite-distance geodesics. We therefore neglect this possibility in the present work and concentrate our attention on potentials in generic infinite-distance limits.

These principal tower product rules were generalized in \cite{Etheredge:2025ahf} to account for heavier towers and branes. These rules come from interpreting the principal plane as being spanned by radions and dilatons. At large volume (i.e., when the radions are taken to be large), the branes may be viewed as either wrapped/unwrapped branes descending from a higher-dimensional theory, KK-modes, or KK-monopoles. Meanwhile, at weak string coupling (i.e., when the dilaton is taken to be large), the brane tensions should scale with integer powers of the string coupling in string units (see \cite{Etheredge:2025ahf} also for a T-duality argument for this). Together, these restrictions lead to rules for brane tensions analogous to the principal tower product rules of \eqref{principaltower}, which we call the radion and dilaton lattice rules, respectively.

\begin{figure}[ht]
		\centering
\includegraphics[width=0.5\textwidth]{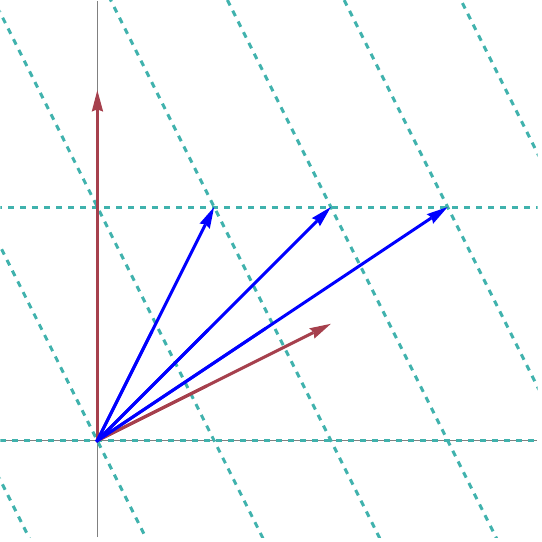}
		\caption{Lattice of $v$-vectors for a 2d principal plane. The principal-tower $\alpha$-vectors are the two burgundian arrows, and the turquoise dashed lines represent loci satisfying the codimension-0 dot lattice conditions with the principal towers. The $v$-vectors must reside at the intersections of the dashed lines, and thus reside on a lattice. The blue arrows are directed at three example $v$-vector lattice sites.}
		\label{figure.lattice}
\end{figure}

To present these rules, we define the generalization of the $\alpha$-vector of a mass to the $\alpha$-vector of a tension using
\begin{align}
    \vec \alpha=-\vec \nabla \log T.
\end{align}
If $\vec \alpha_i$ represents the $\alpha$-vector of a principal tower corresponding to a decompactification from $d$ to $n_i+d$ dimensions, then the $\alpha$-vector of a $(p-1)$-brane in $d$ dimensions satisfies the \emph{\textbf{radion lattice rule}}:
\begin{align}
    \hat \alpha_i\cdot \vec \alpha_p=\frac{p(n_i +d-2)-(p+k_i)(d-2)}{\sqrt{n_i(n_i+d-2)(d-2)}},
    \label{radionpr}
\end{align}
where $n_i=D_{i+1}-D_i$ is the dimension of the $i$th compactification manifold and $k_i$ is an integer that can typically be interpreted as the number of dimensions wrapped by the brane around the $i$th compactification manifold.\footnote{For KK-modes and KK-monopoles, $k_i$ can take more exotic values.} Meanwhile, for a principal tower that is a string oscillator mode, the $\alpha$-vector of a $(p-1)$-brane satisfies the \emph{\textbf{dilaton lattice rule}}:
\begin{align}
    \hat \alpha_\text{osc}\cdot \vec \alpha_p=\frac{p}{\sqrt{d-2}}-\frac{\sqrt{d-2}}2k_\text{osc},
    \label{dilatonpr}
\end{align}
for an integer $k_\text{osc}$ related to the number of $g_s$ factors appearing in the string-unit tension of the $(p-1)$-brane. Together, these rules imply that the $\alpha$-vectors of branes reside on a lattice (see Figure \ref{figure.lattice}), as the number of distinct constraint equations \eqref{radionpr}-\eqref{dilatonpr} for any $\alpha$-vector is equal to the dimension of the principal plane.

Lattice points are labeled by vectors $\vec k$. In a Planckian phase, the $i$th component $k_i$ for non-exotic branes may be viewed as the spacetime dimension of the brane when decompactifying in the direction of the $i$th principal tower. In a stringy phase, the same holds for KK-mode principal towers, but for string oscillator principal towers the component $k_\text{osc}$ is related to the number of string coupling constants that appear in the brane's tension in string units. For a $(p-1)$-brane in a Planckian phase, the position of the $\alpha$-vector is given by \cite{Etheredge:2025ahf}
\begin{align}
    \vec \alpha_p(\vec k)=\sum_i\left(\frac{n_i}{D-2}P-k_i\right)\vec\alpha_i,
\end{align}
where
\begin{align}
    P=p+\sum_ik_i,\qquad D=d+\sum_i n_i\,.
    \label{capitalPD}
\end{align}
In a stringy phase, one of the principal towers is an oscillator mode, and \cite{Etheredge:2025ahf}
\begin{align}
    \vec \alpha_p(\vec k)=\sum_{i\neq \text{osc}}\left(\frac{n_ik_\text{osc}}{2}-k_i\right)\vec \alpha_i+\left(\tilde P-\frac{k_\text{osc}(\tilde D-2)}{2}\right)\vec \alpha_\text{osc},
\end{align}
where
\begin{align}
    \tilde P=p+\sum_{i\neq \text{osc}}k_i\,,\qquad\tilde D=d+\sum_{i\neq\text{osc}}n_i\,.
\end{align}

Another important object of study in the taxonomy program of \cite{Etheredge:2024tok} is the \emph{\textbf{species vector}} $\vec \lambda$, which is defined as the gradient of the logarithm of the species scale \cite{Calderon-Infante:2023ler}:
\be
\vec \lambda=-\nabla \log \Lambda_{\rm QG}\,.
\label{speciesvector}
\ee
The species scale satisfies the dot product rule \cite{Castellano:2023jjt, Etheredge:2024tok}:
\begin{align}
    \vec \lambda\cdot \vec \alpha_\text{lightest}=\frac 1{d-2}\,,
\end{align}
where $\vec \alpha_\text{lightest}$ is the $\alpha$-vector of the lightest principal tower.

In asymptotic regimes of moduli space, the species scale $\Lambda_{\rm QG}$ may be identified with either a string scale or a $D$-dimensional Planck scale associated with the decompactification from $d$ to $D$ dimensions. In the former case, we say that the regime of interest lies in a \textbf{\emph{stringy phase}}. In the latter case, we say that it lies in a \textbf{\emph{Planckian phase}}.
All of the principal towers in a Planckian phase represent KK-modes, and all of the moduli of the principal plane represent radions associated with compactification manifolds. In a stringy phase, one of the principal towers represents a tower of string oscillator modes, and its $\alpha$-vector is equal to the species vector, $\vec \alpha_{\rm osc} = \vec \lambda$. The remaining principal towers represent KK-modes.
In general, a given moduli space will feature both stringy phases and Planckian phases, which are separated by phase transitions.

\section{Principal Potential Terms}\label{sec:LEADING}

In this section, we restrict our attention to several distinguished classes of terms in the asymptotic potential, which we refer to as \emph{\textbf{principal}} terms. In the context of string theory, these principal terms typically represent the leading-order terms in the $g_s$ and $\alpha'$ expansions. We will argue that the dilaton- and radion-dependence of these potential terms precisely matches those of certain codimension-1 or codimension-0 branes. Consequently, the classification of $v$-vectors for these principal potential terms follows immediately from the classification of $\alpha$-vectors for codimension-1 branes and codimension-0 branes.

More precisely, the principal potential terms of interest descend under Kaluza-Klein reduction from fluxes, branes, internal curvature of the compactification manifold, and higher-dimensional potentials. In a critical string theory, an asymptotic potential may also arise through a string-frame cosmological constant (as happens for the $O(16)\times O(16)$ heterotic string). We will now investigate each of these mechanisms in turn.

\subsection{Flux potentials}\label{ss.FLUX}

We begin our analysis by considering asymptotic potentials generated by fluxes threading cycles of a compactification manifold \cite{Taylor:1999ii, Curio:2000sc, Giddings:2001yu}. Specifically, in \S\ref{sss.0form}, we explain how 0-form fluxes generate scalar field potentials of the form $V \sim T_{d-1}^2$, where $T_{d-1}$ is the tension of a codimension-1 brane. In \S\ref{sss.fluxcomp}, we explain how these potentials arise from compactifications. In \S\ref{sss.kinmix}, we discuss kinetic mixing, and in \S\ref{sss.IIKKLT} we consider the example of Type II string theory and the KKLT potential \cite{Kachru:2003aw}.

\subsubsection{0-form flux potentials}\label{sss.0form}

Consider the theory of a 0-form flux coupled to gravity in $d$ dimensions:\footnote{This occurs, for instance, in the Kaloper-Sorbo description of monodromy \cite{Kaloper:2008fb}.} 
\be
S =  \int d^d x \sqrt{-g_d} \left( - \frac{1}{2 } \mathcal{R}_d - \frac{1}{2 g^2 } |F_0|^2 \right) \,.
\label{SF0}
\ee
Here, the flux term $|F_0|^2 \equiv F_{0} \wedge  \star F_0$ acts as a cosmological constant. As one crosses a domain wall, the flux $F_0$ may jump, consistent with Dirac quantization.

More generally, the flux term may depend on the moduli of the theory through the gauge coupling, $g = g(\phi)$. In this case, the flux term acts as a potential for the moduli,
\be
V(\phi) = \frac{1}{2 g^2(\phi)} |F_0|^2\,.
\label{Vphi}
\ee
The flux $F_0$ is dual to a 4-form flux $F_4 \sim \frac{1}{g^2} \star F_0$. Codimension-1 branes, also known as domain walls, are charged under the associated 3-form gauge field, $A_3$. The Weak Gravity Conjecture \cite{Arkanihamed:2006dz} implies that the tension of such a domain wall is bounded above as
\be
T_{d-1}^2 \lesssim \frac{1}{g^2}\,.
\ee
When the taxonomy rules of \S\ref{sec:REVIEW} apply, this bound is approximately saturated:
\be
T_{d-1}^2 \sim \frac{1}{g^2}\,.
\ee
Comparing with \eqref{Vphi}, we have
\be
V \sim T_{d-1}^2\,.
\label{VTdm1}
\ee
That is, the potential scales parametrically with the square of the domain wall tension. This immediately implies a relation between the $v$-vector of the potential term and the $\alpha$-vector of the codimension-1 brane:
\be
\vec v = 2 \vec \alpha_{d-1}\,.
\ee

\subsubsection{Flux potentials from compactification}\label{sss.fluxcomp}

In the context of a Kaluza-Klein compactification,
flux potentials of the form \eqref{Vphi} arise via reduction of a $(P+1)$-form flux on a $(P+1)$-cycle. We begin with a $D$-dimensional action of the form
\be
S_D \supset -\frac{1}{2 e_{P;D}^2 } \int  d^{D}x \sqrt{-g_D} |F_{P+1}|^2\,.
\ee
We now reduce to $d$ dimensions, adding a flux for the $F_{P+1}$ on a $(P+1)$-cycle $\Sigma_{P+1}$. The resulting action contains a potential of the form \eqref{Vphi}, with
\be
g^2 = \text{vol}(\Sigma_{P+1}) e_{P;D}^2 \exp\left(2 \cdot \frac{P(d-2)+D-2}{\sqrt{(D-d)(d-2)(D-2)}} \hat \rho \right)\,,
\ee
where $\rho$ is the canonically normalized radion. Unsurprisingly, this matches the radion-dependence of a codimension-1 brane obtained from reducing a $(D-P-3)$-brane \cite{Etheredge:2022opl, Etheredge:2025ahf}. Consequently, the resulting potential term scales as
\be
 V \sim \frac{1}{g^2} \sim T_{d-1}^2 \sim \exp\left(-2 \cdot \frac{P(d-2)+D-2}{\sqrt{(D-d)(d-2)(D-2)}} \rho \right) \,.
 \label{eqpotflux}
\ee
Hence,
\be
v_\rho  = - \partial_{\rho} \log V = 2 \alpha_{d-1, \rho}^{(D-P-2)} = 2 \cdot \frac{P(d-2)+D-2}{\sqrt{(D-d)(d-2)(D-2)}} \,,
\ee
where $\alpha_{d-1, \rho}^{(D-P-2)}$ is the radion component of the $\alpha$-vector associated with the aforementioned brane.

The connection \eqref{VTdm1} between the flux potential and the tension of a codimension-1 brane ensures that this analysis extends to the case of a multi-step dimensional reduction process, as required for the taxonomy program reviewed in \S\ref{sec:REVIEW} above. In other words, beginning with a flux term $|F_{P_N+1}|^2$ in $D=D_N$ dimensions, we may iteratively reduce to a flux term $|F_{P_i+1}|^2$ in $D_i < D_{i+1}$ dimensions, finally arriving at an $|F_0|^2$ term in $d = D_1$ dimensions. The relation $V \sim \frac{1}{g^2} \sim T_{d-1}^2$ persists even in this more complicated scenario, so
\be
v_{\rho_{i}}  = - \partial_{\rho_i} \log V = 2 \alpha_{\rho_{i}} = 2 \cdot \frac{(D_{i}-P_{i}-2)(D_{i+1}-2)-(D_{i+1}-P_{i+1}-2)(D_{i}-2)}{\sqrt{(D_{i+1}-D_{i})(D_{i}-2)(D_{i+1}-2)}} \,,
\label{genvrho}
\ee
where $\rho_{i}$ is the canonically normalized radion associated with the decompactification from $D_i$ to $D_{i+1}$ dimensions.

In \S\ref{sec:GENERAL}, we will consider this iterative process in greater detail and generality.

\subsubsection{Kinetic mixing}\label{sss.kinmix}

Theories with two or more $P$-form gauge fields generically feature kinetic mixing terms of the form
\be
S_D \subset -\frac{\beta}{2} \int  F_{P+1} \wedge \star H_{P+1}\,,
\label{kinmix}
\ee
where $\beta$ encodes the moduli-dependence of higher-order terms. Upon dimensional reduction, such terms may generate potential terms proportional to $F_0 \wedge \star H_0$, where $F_0$, $H_0$ descend from $F_{P+1}$ and $H_{P+1}$, respectively. We will now argue that the $v$-vectors for these kinetic mixing terms fall into the classification of flux potentials discussed above. That is, they do not produce any novel possibilities for $v$-vectors.

Let us begin by considering the dimensional reduction of a term of the form \eqref{kinmix}. Under such a reduction, these off-diagonal kinetic terms pick up identical radion-dependence to their diagonal cousins, $|F_{P+1}|^2$, $|H_{P+1}|^2$. This means that the radion components of the associated $v$-vector are given simply by \eqref{genvrho}. This means that any novel possibilities must come from the moduli-dependence of the coefficient $\beta$.

However, the Emergent String Conjecture tells us that the $(P+1)$-form fluxes involved in kinetic mixing in an infinite-distance limit should come either from a parent $D$-dimensional theory or from reduction of a higher-form gauge field. The familiar example of the former is the kinetic mixing of the NS-NS 3-form flux $H_3=dB_2$ with the R-R 3-form flux $F_3=dC_2$. However, this kinetic mixing may be repackaged as
\be
S \supset -\frac{1}{2}\int d^{10}x \sqrt{-g_{10}} \left( \frac{1}{e_{2;10}^2} |\tilde F_3|^2 + \frac{1}{e_{10}^2} |H_3|^2  \right)\,,~~~~ \tilde F_3 = F_3 - C_0 H_3\,,
\label{IIBmixing}
\ee
where $e_{2;10}$ and $e_{10}$ are the gauge couplings of $C_2$ and $B_2$, respectively, and $C_0$ is the R-R 0-form. We see that by a suitable choice of field redefinition, the kinetic mixing has been absorbed into the definition of $\tilde F_3$, and we are left with two flux terms of the familiar form \eqref{Vphi}.

Similarly, in the case of Kaluza-Klein reduction, kinetic mixing comes via a dimensional reduction ansatz of the schematic form
\be
F_{P+1}^{(D)} = \omega_{P-p} \wedge F_{p+1}^{(d)} +  B_{P-p} \wedge H_{p+1}\,,
\ee
where $\omega_{P-p}$ is a $(P-p)$-form on the internal space and $B_{P-p}$ is a $(P-p)$-form gauge field in $D$ dimensions whose holonomy $\oint B_{P-p}$ yields an axion $\theta$ in $d$ dimensions. A gauge transformation $\theta \rightarrow \theta + 2 \pi$ then induces $H_{p+1}$ charge from $F_{p+1}$ charge, analogous to the Witten effect \cite{Witten:1979ey}. The gauge kinetic matrix in $d$ dimensions then takes the form
\be
S \supset -\frac{1}{2} \int d^dx \sqrt{-g_d} \left(\frac{1}{e_{H}^2} |H_{p+1}|^2  + \frac{1}{e_{F}^2} |\tilde F_{p+1}|^2\right)\,,~~~~\tilde F_{p+1} = F_{p+1} + \frac{\theta}{2 \pi} H_{p+1} \,.
\ee
A concrete example of this is the kinetic mixing of a 1-form gauge field and a KK-photon on $S^1$ reduction (see e.g. \S3.2.1 of \cite{Heidenreich:2015nta}). The upshot is that, as in \eqref{IIBmixing}, the action is given by a simple sum of flux terms, and the kinetic mixing has been absorbed into the definition of $\tilde F_{p+1}$. Under further dimensional reduction, these two terms will produce two flux terms of the form \eqref{Vphi}. We conclude that---at least in familiar examples of reduction of examples of known string theories---asymptotic potentials involving kinetic mixing fall under the purview of our previous analysis of flux potentials, so they do not generate new possibilities for $v$-vectors.

\subsubsection{Example: Type II string theory and KKLT}\label{sss.IIKKLT}

In a compactification of Type II string theory on a 6-manifold, the R-R gauge coupling $e_{P;10}$ depends on the dilaton. In terms of the canonically normalized 10d dilaton $\hat \phi$, we have
\be
e_{P;10} \sim \exp\left(  \frac{P-4}{\sqrt{8}} \hat \phi\right) \,.
\ee
Upon reduction to 4d, the $(P+1)$-form flux terms generate a 0-form flux potential:
\be
\frac{1}{2 e_{P;10}^2}\int d^{10}x \sqrt{-g_{10}} |F_{P+1}|^2 ~~~\rightarrow ~~~ \frac{1}{2 e_{P;4}^2} \int d^{4}x \sqrt{-g_{4}}  |F_{0}|^2\,.
\ee
This generates a 4d potential of the form
\be
V_{F_{P+1}} \sim \frac{1}{e_{P;4}^2} \sim \exp\left( \frac{4-P}{\sqrt{2}}\hat \phi  - \frac{P+4}{\sqrt{6}} \hat \rho \right)\,.
\label{VFP}
\ee
One can check that this matches with the expression for the R-R flux potential in (3.3) of \cite{Hertzberg:2007wc} (see also (2.19) of \cite{Andriot:2020lea}) after expressing the moduli $\rho$, $\tau$ in terms of the 10d dilaton and 10d radion.\footnote{In particular, note that $\tau$ and $\rho$ in that paper are related to the 4d dilaton and 4d radion, respectively. These are related to the canonically normalized 10d dilaton and 10d radion by a rotation: $\hat \phi_{\rm 4d} = \frac{1}{2} \hat \phi_{\rm 10d} - \frac{\sqrt{3}}{2} \hat \rho_{\rm 10d}$, $\hat \rho_{\rm 4d} = \frac{1}{2} \hat \rho_{\rm 10d} + \frac{\sqrt{3}}{2} \hat \phi_{\rm 10d}$.} Furthermore, the $v$-vector of this potential satisfies $\vec v = 2 \vec \alpha_3^{(D-P-2)}$, where $\vec \alpha_3^{(D-P-2)}$ is the $\alpha$-vector of the 4d 2-brane that descends from a wrapped D$(D-P-3)$-brane in 10d.

Similarly, the NS-NS 2-form gauge coupling scales as
\be
e_{10} \sim \exp\left(\frac{1}{\sqrt{2}} \hat \phi\right)\,.
\ee
Upon reduction, this generates a 0-form flux term:
\be
\frac{1}{2 e_{10}^2} \int d^{10}x \sqrt{-g_{10}}  |H_{3}|^2 ~~~\rightarrow ~~~ \frac{1}{2 e_{4}^2} \int d^{4}x \sqrt{-g_{4}}    |H_{0}|^2\,.
\ee
This yields a potential of the form
\be
V_{H_3} \sim \frac{1}{e_{4}^2} \sim  \exp\left( -\sqrt{2}\hat \phi  - \sqrt{6} \hat \rho \right)\,.
\label{VH3}
\ee
This agrees with (3.3) of \cite{Hertzberg:2007wc}, and the associated $v$-vector satisfies $\vec v = 2 \vec \alpha_3^{\text{NS5}}$, where $\vec \alpha_3^{\text{NS5}}$ is the $\alpha$-vector associated with the 2-brane descending from a wrapped NS5-brane.

A concrete example where these potentials feature prominently is the celebrated KKLT construction \cite{Kachru:2003aw}. Here, the asymptotic potential is generated by fluxes for $H_3$ and $F_3$. These yield an asymptotic potential of the form
\be
V \sim V_{H_3} + V_{F_3}\,,
\ee
where $V_{H_3}$ is given by \eqref{VH3} and $V_{F_3}$ is given by \eqref{VFP} with $P=2$:
\be
V_{F_3} \sim \frac{1}{e_{2;4}^2} \sim \exp\left( \sqrt{2} \hat \phi  - \sqrt{6} \hat \rho \right)\,.
\label{VF3}
\ee
Together, these potential terms stabilize the dilaton $\hat \phi$, so that asymptotically, the effective potential felt by the rolling field takes the form
\be
V_{\rm eff} \sim \exp\left(- \sqrt{6} \hat \rho\right)\,.
\label{Veff}
\ee
This matches the well-known scaling of the KKLT potential \cite{Kachru:2003aw}.

\subsection{Spacetime-filling branes}\label{ss.Spacetimefilling}

A codimension-0 brane of tension $T_d$ contributes to the vacuum energy of a theory in $d$ dimensions. If $T_d=T_d(\phi)$ is moduli-dependent, it generates a potential of the form
\be
V(\phi) \sim T_d(\phi)\,.
\ee
This immediately implies a relation between the $v$-vector of the potential term and the $\alpha$-vector of the codimension-0 brane:
\be
\vec v =  \vec \alpha_{d}\,.
\ee

\subsubsection{Spacetime-filling brane from compactification}\label{sss.spacefillcomp}

One simple way to get a spacetime-filling brane in $d$ dimensions is to start with a $(P-1)$-brane in $D$ dimensions and wrap it on a $(P-d)$-cycle. The resulting $(d-1)$-brane picks up a radion-dependence of the form
\be
V \sim T_d \sim \exp\left( \frac{P(d-2)-d(D-2)}{\sqrt{(D-d)(d-2)(D-2)}} \hat \rho \right)\,.
\label{Tsf}
\ee
The scaling relation $V\sim T_{d}$ applies regardless of the origin of the spacetime-filling brane in question. This means that, as in \S\ref{sss.fluxcomp}, this relation
extends to a multi-step compactification process. See \S\ref{sec:GENERAL} for further discussion.

\subsubsection{Example: Type II string theory}

In 10d string theory, a D$(P-1)$-brane has a tension given by
\be
T_P \sim \exp\left(  \frac{P-4}{\sqrt{8}} \hat \phi\right)\,.
\ee
Compactifying to 4d and wrapping this brane on a $(P-4)$-cycle, we find a spacetime-filling 3-brane that generates a potential of the form
\be
V \sim T_d^{(P)} \sim \exp\left( \frac{P-4}{\sqrt{8}} \hat \phi + \frac{P-16}{ \sqrt{24}} \hat \rho  \right)  \,.
\ee
One can check that this matches with the explicit expression for the potential term in (4.1) of \cite{Hertzberg:2007wc} (see also (2.19) of \cite{Andriot:2020lea}) after expressing the moduli $\rho$, $\tau$ in terms of the canonically normalized 10d dilaton and 10d radion. This immediately implies the relation $\vec v = \vec \alpha_d^{(P)}$, where $\vec \alpha_d^{(P)}$ is the $\alpha$-vector associated with the wrapped D$(P-1)$-brane.

\subsection{Dimensional reduction of a potential}\label{ss.DIMRED}

Suppose a theory in $D$ dimensions has a potential $V_D(\phi)$. After reduction to $d$ dimensions, this yields a $d$-dimensional potential of the form \cite{Rudelius:2022gbz}:
\be
V_d(\phi, \hat \rho) \sim V_D(\phi) \exp\left( - 2\sqrt{\frac{D-d}{(D-2)(d-2)}} \hat \rho \right)\,.
\label{VdVD}
\ee
This scaling matches that of \eqref{eqpotflux} with $P = -1$. This corresponds to a flux potential $V_D \sim|F_0|^2$ in $D$ dimensions, which scales with the square of the tension of a codimension-1 brane, $V_D  \sim T_{D-1}^2$. The agreement between \eqref{VdVD} and \eqref{eqpotflux} with $P = -1$ shows that this relation is preserved under dimensional reduction, $V_d  \sim T_{d-1}^2$, where $T_{d-1}$ is the tension of the codimension-1 brane in $d$ dimensions that comes from wrapping the codimension-1 brane in $D$ dimensions around the compactification manifold.

The scaling \eqref{VdVD} also matches \eqref{Tsf} with $P=D$. This corresponds to wrapping a spacetime-filling brane of tension $T_D$ to produce a spacetime-filling brane of tension $T_d$. The match between \eqref{VdVD} and \eqref{Tsf} shows that the relation $V_D \sim T_D$ is preserved under dimensional reduction, $V_d \sim T_d$.

In the case of a flux potential or spacetime-filling brane studied in \S\ref{ss.FLUX}-\S\ref{ss.Spacetimefilling}, there are physical branes whose tensions $T_{D-1}$, $T_D$ match the scaling behavior of the potential, $V_D \sim T_{D-1}^2$, $V_D \sim T_D$. Such branes may be wrapped to produce the relevant codimension-1/codimension-0 branes in $d$ dimensions. Below, we will encounter examples in which there is not necessarily a physical brane of codimension-1 or codimension-0 associated with the potential $V_D$, and consequently there is not necessarily a physical brane associated with $V_d$ in $d$ dimensions. For our purposes, however, the crucial point is simply that the associated $v$-vector in $d$ dimensions is consistent with the taxonomy rules for codimension-1 branes and/or codimension-0 branes (i.e., $\vec v^{(d)} = 2 \vec \alpha^{(d)}_{d-1}$ or $\vec v = \vec \alpha^{(d)}_{d}$) provided that the $D$-dimensional $v$-vector is consistent with the respective taxonomy rules for codimension-1 and/or codimension-0 branes in $D$ dimensions (i.e., $\vec v^{(D)} = 2 \vec \alpha^{(D)}_{D-1}$ or $\vec v = \vec \alpha^{(D)}_{D}$).

\subsection{Reduction with internal curvature}\label{ss.Internalcurvature}

Compactifying from $D=d+n$ to $d$ dimensions on an internal $n$-dimensional manifold with nonzero curvature $\mathcal{R}_n$ yields a potential of the form \cite{Andriot:2020lea,Rudelius:2022gbz}
\be
V \sim - \mathcal{R}_n \exp\left(- 2 \sqrt{\frac{D-2}{(D-d)(d-2)}} \hat \rho \right) \,.
\label{VIC}
\ee
Setting $P=0$ in \eqref{eqpotflux}, we see that this precisely matches the scaling of an $|F_1|^2$ flux term in $D$ dimensions that is reduced to an $|F_0|^2$ flux in $d$ dimensions. Thus, from the perspective of our taxonomy program, a potential generated by internal curvature is equivalent to a flux potential generated by the reduction $|F_1|^2 \rightarrow |F_0|^2$, as studied in \S\ref{ss.FLUX}. This means that the potential generated by internal curvature scales precisely like the square of the tension of a codimension-1 brane that descends from a codimension-2 brane in the parent $D$-dimensional theory, and the $v$-vector is related to the $\alpha$-vector of this brane via
\be
v_\rho = 2\alpha_{d-1, \rho}^{(D-2)} \,.
\label{valphaIC}
\ee

In contrast to the flux potentials studied in \S\ref{ss.FLUX}, however, there does not generically seem to be any physical codimension-1 brane associated with this potential. One exception to this comes in the case of a $T^3$ compactification of Type II string theory, which features codimension-1 branes that come from the KK monopole associated to the circle reduction from 10d to 9d.\footnote{A KK monopole is codimension-3 in general. Here, a 9d KK monopole descends to a codimension-1 brane in 7d.} A straightforward computation reveals that the tension of this codimension-1 brane scales with the canonically normalized, overall volume modulus of the $T^3$ as
\be
T_{d-1} \sim \exp\left(- \sqrt{\frac{8}{15}} \hat \rho \right)\,.
\ee
Setting $V \sim T_{d-1}^2$, this precisely matches \eqref{VIC} with $D=10$, $d=7$.

Even in this case, however, there is no codimension-2 brane in $D=10$ dimensions associated with the potential; the codimension-1 brane of interest has a geometric origin. In general, then, the relation \eqref{valphaIC} should be viewed as a formal mathematical correspondence between $v$-vectors and $\alpha$-vectors rather than a physical relation between potentials and brane tensions.

 Remarkably, this formal correspondence extends to a multi-step compactification process, though this requires further justification. For simplicity, let us consider a two-step compactification from $D=D_3$ dimensions to $D_2$ dimensions to $d=D_1$ dimensions. We then have two options to generate a $d$-dimensional potential from internal curvature:
 \begin{enumerate}
     \item We may first reduce on a manifold with nonzero curvature to generate a potential in $D_2$ dimensions, which descends directly to a potential in $d=D_1$ dimensions.
     \item We may induce a potential in $d=D_1$ dimensions from compactifying the theory in $D_2$ dimensions on a manifold with nonzero curvature.  
 \end{enumerate}
Of course, if both compactification manifolds have nonzero curvature, we may simultaneously generate two potential terms, one from each of these processes.
 
 In the first case, \eqref{valphaIC} implies that in the first reduction from $D$ to $D_2$ dimensions, the radion-dependence of the potential matches that obtained by reducing an $|F_1|^2$ flux term to a $|F_0|^2$ flux term.
  Furthermore, the analysis of \ref{ss.DIMRED} ensures that in the compactification from $D_2$ to $D_1$ dimensions, the potential term acquires a radion-dependence matching that of an $|F_0|^2$ flux term reduced to an $|F_0|^2$ flux term. Using the correspondence between flux terms and branes discussed in \S\ref{ss.FLUX}, we conclude that
 \be
\vec{v}  = 2 \vec \alpha_{d-1} \,,
 \ee
where $\vec{\alpha}_{d-1}$ represents the $\alpha$-vector associated with a codimension-2 brane in $D=D_3$ dimensions that is reduced to give a codimension-1 brane in $D_2$ dimensions and further wrapped to give a codimension-1 brane in $D_3$ dimensions.

The second case requires a novel computation. Here, the potential in $d$ dimensions is generated by the 
Einstein-Hilbert term in $D_2$ dimensions, which descends directly from the Einstein-Hilbert term in $D$ dimensions and (in Einstein frame) does not depend on the radion $\rho_2$. Similarly, in a reduction of the form
\be
 \frac{1}{e_{0;D_3}^2} \int d^{D_3}x \sqrt{-g_{D_3}} |F_1|^2 ~~~~ \rightarrow  \frac{1}{e_{0;D_2}^2} \int d^{D_2}x \sqrt{-g_{D_2}} |F_1|^2\,,
\ee
the formula \eqref{genvrho} with $D_i=D_2$, $D_{i+1}= D_3$, $P_i=P_{i+1}=0$ shows that the gauge coupling $e_{0;D_2}^2$ does not depend on the radion, i.e., $\alpha_{\rho_2} = 0$.

Combining this result with \eqref{valphaIC}, we again conclude that the radion-dependence of the potential term in $d$ dimensions matches that of a flux potential term $|F_0|^2$ that descends from an $|F_1|^2$ term in $D_2$ dimensions, which in turn descends from an $|F_1|^2$ term in $D$ dimensions. By the correspondence between flux terms and branes discussed in \S\ref{ss.FLUX}, we again conclude that $\vec{v}  = 2 \vec \alpha_{d-1}$,
where here $\vec{\alpha}_{d-1}$ represents the $\alpha$-vector associated with a codimension-2 brane in $D=D_3$ dimensions that is wrapped to give a codimension-2 brane in $D_2$ dimensions and further reduced to give a codimension-1 brane in $D_1=d$ dimensions.

This conclusion generalizes immediately to compactification processes involving more than two steps: the $v$-vector associated with a potential in $d$ dimensions obtained by internal curvature of some compactification manifold is equal to twice the $\alpha$-vector of a codimension-1 brane that descends from a codimension-2 brane in the original theory in $D$ dimensions, $\vec v = 2 \vec \alpha_{d-1}^{(D-2)}$.

One important caveat is that the (Einstein-frame) Einstein-Hilbert term has trivial dilaton dependence in the parent $D$-dimensional theory, hence any potential obtained from internal curvature does not depend on the $D$-dimensional dilaton. This is in contrast to the case of the $|F_1|^2$ term in Type IIB string theory, which does depend on the dilaton.

\subsection{String-frame cosmological constant}\label{ss.SFCC}

The $O(16) \times O(16)$ heterotic string in 10d obtains a potential at 1-loop, which takes the form of a cosmological constant in the string frame. More generally, a $D$-dimensional cosmological constant in the string frame picks up dilaton dependence after Weyl rescaling:
\be
V \sim \exp\left(\frac{D}{\sqrt{D-2}} \hat \phi\right)\,.
\label{stringframecc}
\ee
Comparing with \eqref{dilatonpr}, we see that this matches the scaling behavior of a codimension-1 brane in $D$ dimensions with $P=D$, $V(\phi) \sim T_{D-1}^2(\phi) \Rightarrow v_\phi =  2 \alpha_{D-1,\phi} = -D/\sqrt{D-2}$.

In the context of massive Type IIA string theory, this particular $\alpha$-vector can be identified with a physical codimension-1 brane.
Here, the Romans mass term takes the form
\be
-\int d^{10}x   \sqrt{-g_{10}} \frac{1}{2 e_{0;10}^2} |F_0|^2\,,
\ee
which yields a potential
\be
V \sim T_9^2 \sim \exp\left(\frac{10}{\sqrt{8}} \hat \phi \right)\,,
\ee
where $T_9$ is the tension of a D8-brane. 
This potential agrees precisely with \eqref{stringframecc}, which is related to the fact that $|F_0|^2$ and a string-frame cosmological constant are both independent of the dilaton in string frame and pick up identical factors from Weyl rescaling when moving to Einstein frame.\footnote{In particular, the only dilaton dependence comes from the transformation of $\sqrt{-g_D}$ under Weyl rescaling.}

More generally, however, the potential \eqref{stringframecc} does not necessarily correspond to the tension of a physical brane in $D$ dimensions. For our purposes, the key point is simply that the $\alpha$-vector $\alpha_{D-1,\phi}$ is consistent with the taxonomy rules for codimension-1 branes in $D$ dimensions.

As we saw in \S\ref{ss.DIMRED}, this correspondence persists under dimensional reduction: $V_D \sim T_{D-1}^2$ in $D$ dimensions implies $V_d \sim T_{d-1}^2$ in $d$ dimensions, where $T_{d-1}$ is the tension of the codimension-1 brane obtained by wrapping the codimension-1 brane in $D$ dimensions around the compactification manifold. Thus, as in the case of flux potentials and internal curvature studied above, the $v$-vector associated with a string-frame cosmological constant satisfies $\vec v = 2 \vec \alpha_{d-1}$, where $\vec \alpha_{d-1}$ is consistent with the taxonomy of codimension-1 branes.

\subsection{Axion potentials}\label{ss.axionpot}

Axion potentials, generated by instanton effects, represent another prominent class of asymptotic potentials. These take the schematic form
\be
V(\theta, \tau) \sim V_0 \exp(- 2 \pi \tau) \cos(\theta)\,,
\ee
where $\theta$ is a compact axion and $\tau$ is a noncompact saxion. Such potentials play an important role in the context of axion physics and the KKLT construction of (A)dS vacua \cite{Kachru:2003aw}, but they are not so relevant for our purposes of studying the \emph{asymptotic} potential. The reason for this is that, in asymptotic regimes, the kinetic term of the saxion $\tau$ picks up a logarithmic divergence:
\be
\mathcal{L} \supset -\frac{1}{2 a^2 \tau^2} (\nabla \tau)^2\,. 
\ee
As a result, the potential takes the form of a double exponential when expressed in terms of the canonically-normalized saxion $\hat \tau$:
\be
V \sim \exp\left(- 2 \pi e^{a \hat \tau} \right) \cos(\theta)\,.
\ee
These doubly exponentially suppressed terms will be dominated by any exponentially suppressed term in the asymptotic potential. Relatedly, the $v$-vector associated to such a term diverges in the limit $\hat \tau \rightarrow \infty$:
\be
v_{\hat \tau} = -\partial_{\hat \tau} \log V = 2 \pi a e^{a \hat \tau} ~\rightarrow ~ \infty \,.
\ee
It would be interesting to explore a taxonomy of such potentials, perhaps through a doubly logarithmic version of a $v$-vector, $\nabla \log(-\log V)$. We leave this for future work, however, and in this paper we restrict our analysis to asymptotic potentials with finite $v$-vectors.

\section{General Potential Terms}\label{sec:GENERAL}

In general, the taxonomic program developed in \cite{Etheredge:2024tok, Etheredge:2025ahf} and reviewed in \S\ref{sec:REVIEW} above suggests that $\alpha$-vectors of particles and branes and $v$-vectors of potential terms can be obtained through an iterative process of compactification from higher dimensions to lower dimensions:
\be
D=D_N \rightarrow D_{N-1} \rightarrow ... \rightarrow D_1 = d\,.
\label{iterativecomp}
\ee
where $D_i < D_{i+1}$. To classify the $v$-vectors of the $d$-dimensional theory, we want to understand the paths that a given term in the $D$-dimensional theory might take in the iterative compactification process to produce a potential term in the $d$-dimensional theory.

To begin, we focus our attention on terms in the $D$-dimensional action, including the Einstein-Hilbert term $\mathcal{R}_D$ and flux terms $|F_{P+1}|^2$ studied in \S\ref{sec:LEADING}.
We will see that the $v$-vectors of these terms are determined uniquely by their weights under Weyl rescaling (henceforth referred to as \emph{\textbf{Weyl weights}}) and, in the case of a stringy phase, by their (string-frame) scaling with the string coupling $g_s$.\footnote{Weyl weights have appeared previously in the study of the Distance Conjecture \cite{Lust:2019zwm, Kehagias:2019akr}.}
The quantization of Weyl weights and powers of $g_s$ in string perturbation theory ensures that the associated $v$-vectors are also quantized; that is, they reside in a lattice. We extend this to the case of potentials generated by spacetime-filling branes.

\subsection{Weyl weights and compactifications}\label{ss.weylweights}

Under a Weyl transformation of the metric,
\be
g_{MN} \rightarrow \Omega(x)^2 g_{MN} \,,~~~\Omega(x)=e^{\omega(x)}\,,
\ee
a term $\mathcal{O}$ in the Lagrangian of a $D$-dimensional theory will generically transform as
\be
\mathcal{O} ~\rightarrow ~ \Omega^{-w(\mathcal{O})} \mathcal{O} + ...\,,
\ee
where $w(\mathcal{O})$ is the Weyl weight of $\mathcal{O}$ and $...$ indicates $\mathcal{O}$-independent terms that are irrelevant for our purposes.
For example, a flux term transforms as 
\be
|F_{P+1}|^2 ~ \rightarrow ~\Omega^{-2(P+1)} |F_{P+1}|^2 \,,
\ee
so it has Weyl weight $w = 2(P+1)$.
Meanwhile the Ricci scalar transforms as
\be
\mathcal{R}_D ~\rightarrow ~\Omega^{-2}
\left(\mathcal{R}_D - 2(D-1) \nabla^2 \omega - (D-2)(D-1) (\nabla \omega)^2 \right)\,,
\ee
so it has Weyl weight 2. More generally, the Weyl weight of a term comprised of fluxes and/or curvatures ($R^2$, $R F^2$, $F^4$, etc.) will be a non-negative, even integer: $w \in 2 \mathbb{Z}_{\geq 0}$. Morally, $\frac{1}{2}w$ counts the number of factors of the inverse metric $g^{\mu \nu}$ appearing in the relevant operator. Note that under a Weyl transformation, the action will also pick up an overall factor of $\Omega^{D}$ from the transformation of the measure, $\sqrt{-g_D} \rightarrow \Omega^D \sqrt{-g_D}$.

Under compactification, a single term in $D$ dimensions may produce multiple terms of different Weyl weights in $d$ dimensions. For example, under reduction to $d$ dimensions, a flux term $|F_{P+1}^{(D)}|^2$ in $D$ dimensions may give rise to both a $|F_{P+1}^{(d)}|^2$ term and a $|F_{p+1}^{(d)}|^2$ term, where $p < P$. The Weyl weights of these respective terms are given by
\be
w(|F_{P+1}^{(d)}|^2) = w(|F_{P+1}^{(D)}|^2) = 2 (P+1)\,,~~~~ w(|F_{p+1}^{(d)}|^2) = 2 (p+1)\,.
\label{wFluxexample}
\ee

Given a sequence of compactifications from $D=D_N$ dimensions to $D_1=d$ dimensions, 
we may thus assign to each term in the action a vector $\vec w \in  \mathbb{Z}^N$, where $w_i \geq 0$ represents the Weyl weight of the associated term in the action in $D_i$ dimensions. For example, in the simple $N=2$ compactification considered in \eqref{wFluxexample}, the $w$-vectors associated with the flux terms $|F_{P+1}^{(d)}|^2$, $|F_{p+1}^{(d)}|^2$ are given respectively by
\be
\vec{w}(|F_{P+1}^{(d)}|^2) = 2(P+1,P+1)\,,~~~~ \vec{w}(|F_{p+1}^{(d)}|^2) = 2(p+1,P+1) \,.
\ee

Any term in the potential $V(\phi)$ necessarily has vanishing Weyl weight, $w(V(\phi)) = 0$. Therefore, we must have
\be
w_1 = 0\,,
\ee
for any term in the potential.
Typically, we expect that $w_i$ will increase with increasing $i$. However, there is one notable exception to this: the kinetic term $|F_2|^2$ of a Kaluza-Klein photon of $D_i$ dimensions has $w_i=4$, yet it descends from the Einstein-Hilbert term $\mathcal{R}$ in $D_{i+1}$ dimensions, which has $w_{i+1}=2$.

\subsection{Dilaton-dependence}\label{ss.dd}

In \S\ref{ss.vfromw}, we will show that a pair of consecutive Weyl weights $w_i$, $w_{i+1}$ uniquely determines the component of the $v$-vector associated with the decompactification from $D_i$ to $D_{i+1}$ dimensions. This immediately  implies that there is a 1-1 correspondence between $w$-vectors and $v$-vectors in any Planckian phase, since all components of the $v$-vector in such a phase correspond to decompactifications.

In the case of a stringy phase, however, we must also account for the dilaton component of the $v$-vector. To this end, let us consider a string theory in $D$ dimensions, which features a dilaton $\Phi$ and a string coupling constant $g_s = e^\Phi$.\footnote{Here, the conventionally normalized dilaton $\Phi$ is related to the canonically normalized dilaton $\phi$ by $\Phi = \frac{\sqrt{D-2}}{2} \phi $.} The low-energy effective field theory then admits a expansion in the parameter $g_s$:
\be
\mathcal{L} =  \sum_{g=0}^\infty \sum_{n=0}^\infty \sum_{\mathcal O} g_s^{2 g +n - 2} c_{\mathcal O} \mathcal{O}_{g,n}\,,
\label{gsexpansion}
\ee
where $g$ represents the genus of the string worldsheet and $n$ represents the number of punctures associated with the operator $\mathcal{O}_{g,n}$.

For a given operator $\mathcal{O}_{g,n}$, let us define the integral quantity $k_g(\mathcal{O})$ as
\be
k_g(\mathcal{O}) = 2g + n -2\,,
\label{kgdef}
\ee
so that $k_g \geq -2$ measures the scaling of the term with the string coupling $g_s$.

The $g_s$ expansion in \eqref{gsexpansion} is most naturally understood in the string frame. To determine the $v$-vector associated with a given term, however, we must consider the dilaton-dependence of the term in the Einstein frame. Fortunately, as we will briefly review in \S\ref{ss.vfromw}, the relation between the two is uniquely determined by the Weyl weight of the term in question. We thus conclude that, within a stringy phase, the dilaton-component of a $v$-vector is uniquely determined by the $D$-dimensional Weyl weight $w_N$ and the integer $k_g$.

\subsection{$v$-vectors from $w$-vectors}\label{ss.vfromw}

In this subsection, we establish the results claimed in \S\ref{ss.weylweights} and \S\ref{ss.dd}: the $v$-vector of a potential term in $d$ dimensions associated with a term in the $D$-dimensional action is uniquely determined by its $w$-vector of Weyl weights $\vec{w}$ and, in the case of a stringy phase, by its $g_s$-scaling coefficient $k_g$.

\subsubsection{Radion components}

Let us first consider the familiar reduction ansatz from $D$ to $d=D-n$ dimensions:
\be
ds_{D}^2 =  \Omega(x)^{-2n/(d-2)} ds_{d}^2(x) + \Omega^{2}(x) ds_{n}^2(y)\,,
\label{dRansatz}
\ee
where in terms of the canonically normalized radion $\rho$, we have
\be
\Omega = \exp\left(\sqrt{\frac{d-2}{n(D-2)}} \rho \right) \, .
\label{Omegavsrho}
\ee
Note that the factor $\Omega(x)^{-2n/(d-2)}$ essentially acts as Weyl transformation on the $d$-dimensional metric $ds_d^2$, while the factor $\Omega(x)^{2}$ acts as a Weyl transformation on the $n$-dimensional metric $ds_n^2$. 

Now, let us consider the reduction of an operator of Weyl weight $w_D$ in $D$ dimensions to produce an operator of Weyl weight $w_d$ in $d$ dimensions.
Plugging in the ansatz \eqref{dRansatz}, we see that we get a factor of $\Omega^{-\frac{nd}{d-2} + n}$ from the $\sqrt{-g}$ in the action, a factor of $\Omega^{\frac{n w_d}{d-2}}$ 
associated with the transformation of $ds_d^2$,
and a factor of $\Omega^{w_d-w_D}$ associated with the Weyl transformation of $ds_n^2$.
This yields an overall factor of
\be
\Omega^{-\frac{nd}{d-2} + n} \cdot \Omega^{\frac{n w_d}{d-2}} \cdot \Omega^{w_d-w_D} = \Omega^{\frac{n(w_d-2)}{d-2} + w_d-w_D}\,.
\ee

Plugging in \eqref{Omegavsrho}, we find that the $d$-dimensional term in the action scales as
\be
\mathcal{O}_d \sim \exp\left( \sqrt{\frac{d-2}{n(D-2)}} \left( \frac{ n (w_d-2)}{d-2} + w_d-w_D \right) \rho \right)\,.
\ee
This yields a $v$-vector of
\be
v_\rho = \sqrt{\frac{d-2}{n(D-2)}} \left( -\frac{n(w_d-2)}{d-2} +w_D-w_d \right)\,.
\ee

This formula generalizes straightforwardly. Given a $w$-vector $\vec w$ and a corresponding sequence of dimensions $(d=D_1, D_2, ..., D_N=D)$, we find 
\begin{align}
v_{\rho_{i}}&=  \sqrt{\frac{D_{i}-2}{(D_{i+1}-D_{i})(D_{i+1}-2)}} \left( -\frac{ (D_{i+1}-D_{i}) (w_{i}-2)}{D_{i}-2} + w_{i+1} - w_{i} \right) \nonumber \\
&= \sqrt{\frac{1}{(D_{i+1}-D_{i})(D_{i+1}-2)(D_{i}-2)}}\Big( (w_{i+1} - 2)(D_{i}-2) - (w_{i}-2)(D_{i+1}-2) \Big)
\,.
\label{vrhoi}
\end{align}
This is the desired formula for the radion-components of the $v$-vector in terms of the components of the $w$-vector. Note that the Weyl weight $w_i$ is always a (non-negative) integer, which forces the $v$-vector to reside on a lattice.

\subsubsection{Dilaton component}

In the case of a stringy phase, we must also consider the dilatonic component of the $v$-vector. We claim that this depends exclusively on the Weyl weight $w_D=w_N$ of the theory in $D=D_N$ dimensions as well as the $g_s$-scaling coefficient $k_g$ defined in \eqref{kgdef}.

In terms of the canonically normalized dilaton $\phi$, we may write
\be
g_s = \exp \left(\frac{\sqrt{D-2}}{2} \phi\right) \,.
\ee
This implies that a term with $g_s$-scaling $k_g$ comes with a factor of $\exp ( k_g \frac{\sqrt{D-2}}{2} \phi)$ in string frame.

To move from string frame to Einstein frame, we perform a Weyl rescaling
\be
g_{\mu \nu} \rightarrow \Omega^2 g_{\mu \nu}\,,~~~~ \Omega = \exp\left(\frac{1}{\sqrt{D-2}}  \phi\right)\,.
\ee
This Weyl rescaling gives a factor of $\Omega^D$ associated with the transformation of the measure $\sqrt{-g_D}$ as well as a factor of $\Omega^{- w_D}$ due to the Weyl weight of the operator. In total, then, we find that in Einstein frame, the term in question scales as
\be
\mathcal{O} \sim \exp\left(  \frac{1}{\sqrt{D-2}} \left( \frac{k_g (D-2)}{2} + D -  w_D \right)  \phi \right)  \,.
\ee
Hence we have
\be
 v_\phi  =  \frac{1}{\sqrt{D-2}} \left( -\frac{k_g (D-2)}{2} - D +  w_D \right)    \,.
 \label{vphi}
\ee
This is the desired formula, which relates the dilaton-component of the $v$-vector to the $D$-dimensional Weyl weight $w_D=w_N$ and the $g_s$-scaling coefficient $k_g$. Once again, $k_g$ and $w_D$ are both integers, so $v_\phi$ is quantized.

\subsubsection{Example: KKLT and LVS}

The asymptotic behavior of KKLT potential was discussed above in \S\ref{sss.IIKKLT}. The leading terms come from the Calabi-Yau compactification of the $|H_3|^2$ and $|\tilde F_3|^2$ fluxes of Type IIB string theory. In the language of the present section, these have respective $w$-vectors and $g_s$-scaling coefficients
\begin{subequations}
\begin{align}
w_{H} = (0,6)\,,~~~~ k_{g,H} =  -2\,, \\
w_{F} = (0,6)\,,~~~~ k_{g,F} =  0\,.
\end{align}
\end{subequations}
From \eqref{vrhoi}, \eqref{vphi}, these yield respective $v$-vectors
\be
v_H = (\sqrt{6}, \sqrt{2})\,,~~~~ v_F = (\sqrt{6}, -\sqrt{2}) \,.
\ee
These indeed match the scaling behavior of the potential terms in \eqref{VH3} and \eqref{VF3}. Together, these terms stabilize the dilaton, producing an effective, asymptotic potential of the form \eqref{Veff}. 

In the Large Volume Scenario (LVS) \cite{Balasubramanian:2005zx}, the asymptotic potential is given at leading order by a term of the form
\be
V \sim \frac{\xi }{\mathcal{V}^3} |W|^2 \sim \exp\left(- 3 \sqrt{\frac{3}{2}} \rho\right)\,,
\label{VLVS}
\ee
where $\xi$ is a constant, $W$ is the flux superpotential, $\rho$ is the canonically normalized radion, and $\mathcal{V}$ is the volume of the Calabi-Yau manifold $X$ in string units.
The flux superpotential $W$ is given by the integral \cite{Gukov:1999ya}:
\be
W = \int_X G_3 \wedge \Omega\,,
\ee
where $\Omega$ is the holomorphic $(3,0)$-form of $X$ and $G_3 = F_3 - \tau H_3$ is a complexified version of the modified 3-form $\tilde F_3$. This potential term descends from a term in the 10d action of the schematic form
\be
\mathcal{L} \supset |G_3|^2 \mathcal{R}^3\,.
\ee
This term has Weyl weight $w = 12$, so the $w$-vector associated with this potential term is given by
\be
\vec w = (0, 12)\,.
\ee
The dilaton is stabilized, so it does not play a role in the asymptotic potential, but the radion-component of the $v$-vector is determined by \eqref{vrhoi} to be
\be
v_{\rho} = 3\sqrt{\frac{3}{2}}\,,
\ee
which matches \eqref{VLVS}.

Note that the KKLT and LVS constructions of (A)dS vacua in string theory rely on a cancellation of terms. Such a cancellation necessarily occurs in the interior of moduli space, where our taxonomy rules break down and subleading terms in the asymptotic potential compete with leading terms. It would be worthwhile to see if our taxonomy program produces any restrictions on the potential in the interior of moduli space, but this lies beyond the scope of the present work.

\subsection{Spacetime-filling branes, again}

In \S\ref{ss.vfromw}, the formulas \eqref{vrhoi} and \eqref{vphi} determine the $v$-vector uniquely in terms of the $w$-vector of Weyl weights $\vec w$ and the $g_s$-scaling coefficient $k_g$. In \S\ref{ss.weylweights}-\S\ref{ss.dd}, the quantities $\vec w$, $k_g$ were defined as properties of a term in the Lagrangian. However, it turns out that these quantities---as well as the formulas \eqref{vrhoi}, \eqref{vphi}---apply also to potentials generated by spacetime-filling branes.

Let us once again consider a sequence of compactifications from $D=D_N$ dimensions to $d=d_1$ dimensions. Consider a $(P_N-1)$-brane in the parent $D_N$-dimensional theory, which is iteratively wrapped around cycles of the compactification manifold to produce a $(P_i-1)$-brane in $D_i$ dimensions. As discussed in \S\ref{ss.Spacetimefilling}, a potential in $d$ dimensions is associated with the tension of a spacetime-filling brane, i.e., we have $V \sim T_d$ provided $P_1 = d$.

As a result, the radion components of the $v$-vector associated with this potential term are given simply by the $\alpha$-vector components associated with a $(P_{i}-1)$-brane in $D_i$ dimensions that descends from a $(P_{i+1}-1)$-brane in $D_{i+1}$ dimensions \cite{Etheredge:2022opl}:
\be
v_{\rho_i} = \alpha_{\rho_i} =  \frac{1 }{\sqrt{(D_{i+1} - D_i)(D_{i+1}-2)(D_i-2)}}  \left( P_i (D_{i+1}-2)- P_{i+1} (D_i - 2) \right) \,.
\ee
Comparing with \eqref{vrhoi}, we see that this matches the $v$-vector derived from a $w$-vector $\vec{w}$ if we set
\be
w_{i} = D_i - P_i\,.
\ee
Thus, we may associate a $w$-vector to any brane, with $w_i$ given simply by the codimension of the brane in $D_i$ dimensions. As above, a potential term requires a $w$-vector with $w_1=0$, which implies that the brane in question is codimension-0 (i.e., spacetime-filling).

In the case of a Planckian phase, then, this $w$-vector uniquely determines the associated $v$-vector. In the case of a stringy phase, an extra integer $k_g$ is needed to describe the $g_s$-dependence of the brane via \eqref{vphi}. Here, $k_g$ measures the scaling of the $(P_N-1)$-brane tension with $g_s$ in $D$-dimensional string units:
\be
T_{P_N} \sim g_s^{k_g} M_{\rm string}^{P_N}\,.
\ee
For example, the tension of a D$(P-1)$-brane in 10d Type II string theory is \cite{Polchinski:1998rr}
\be
T_P = \frac{1}{g_s (2 \pi)^{P-1} (\alpha')^{\frac{P}{2}} } \sim g_s^{-1} \,
\ee
hence $k_g=-1$. The codimension of this brane is given by $w_D = D-P =10-P$. Plugging these values into the formula \eqref{vphi} yields 
\be
v_\phi = \frac{4-P}{\sqrt{8}}\,,
\ee
which correctly reproduces the dilaton-dependence of the D$(P-1)$-brane (see e.g. Table 1 of \cite{Etheredge:2024amg}).

Similarly, the NS5-brane of Type II string theory has
\be
T_{\rm NS5} = \frac{1}{g_s^2 (2 \pi)^{5} (\alpha')^{3}} \sim g_s^{-2} \,
\label{TNS5}
\ee
hence $k_g=-2$. Setting $w_D = 10-6=4$ equal to the codimension of the NS5-brane, \eqref{vphi} yields 
\be
v_\phi = \frac{2}{\sqrt{8}} = \frac{1}{\sqrt{2}}\,,
\ee
which correctly reproduces the dilaton-dependence of the NS5-brane \cite{Etheredge:2024amg}.

We conclude that the formalism of \S\ref{ss.vfromw} extends to the case of spacetime-filling branes provided we set $w_i$ to be the codimension of the brane in $D_i$ dimensions and set $k_g$ to measure the $g_s$-dependence of the brane tension in string units. Once again, both $w_i$ and $k_g$ are integrally quantized, and $w_i$ is necessarily non-negative: $w_i \in \mathbb Z_{\geq 0}$, $k_g \in \mathbb Z$. 

Note, however, that while the Weyl weights $w_i$ of terms comprised of curvatures and fluxes were always even integers, the codimension of a brane $w_i$ may be either even or odd. This means that, in general, the lattice of $v$-vectors will be generated by $\alpha$-vectors of codimension-0 branes.

\subsection{Casimir Energies}
\label{ss.Casimir}

Casimir energies represent another source of potential terms. These energies are associated with matter fields with different possible boundary conditions around the compactification manifold, and in $d$ dimensions, they scale with the radius of compactification $R$ as \cite{Arkani-Hamed:2007ryu}
\be
V_{\text{Csmr}} \sim \frac{1}{R^d} \exp(-2 \pi m R)\,, 
\label{Casimirgeneral}
\ee
where $m$ is the mass of the particle in question and the overall sign depends on the boundary conditions and spin of the particle. The radius $R$ scales exponentially with the canonically normalized radion, so for a massive particle, this potential is doubly exponentially suppressed. We ignore such potentials in the present work; see \S\ref{ss.axionpot} above.

For massless particles, however, the Casimir energy \eqref{Casimirgeneral} decays with the canonically normalized radion $\rho$ as
\be
V_{\text{Csmr}} \sim \exp\left(- d \sqrt{\frac{D-2}{(D-d)(d-2)}} \rho\right)
\label{Casimirrho}
\ee
in a compactification from $D$ to $d$ dimensions, where we have accounted for the factor associated with the transformation to $d$-dimensional Einstein frame. This potential fits squarely into the scope of our taxonomy program.

Let us therefore, as usual, consider a sequence of decompactifications from $d=D_1$ to $D=D_N$ dimensions. A Casimir potential term associated with the $i$th decompactification manifold (from $D_i$ to $D_{i+1}$ dimensions) will then have a $v$-vector given by
\be
v_{\rho_i}^{\text{Csmr}} = \begin{cases}
        2 \sqrt{\frac{D_{i+1}-D_i}{(D_{i+1}-2)(D_{i}-2)}} & \text{if $j < i$}  \\
            D_i \sqrt{\frac{D_{i+1}-2}{(D_{i+1}-D_i)(D_{i}-2)}} & \text{if $j = i$} \\
           0 &  \text{if $j > i$}\,.
		 \end{cases} 
\ee
Here, the case $i=j$ corresponds to the introduction of the Casimir potential \eqref{Casimirrho} associated with the compactification from $D_{i+1}$ to $D_i$ dimensions, $j < i$ corresponds to the familiar case of the dimensional reduction of a potential, and $j > i$ corresponds to the fact that the Casimir energy depends only on the radion and not on other moduli of the theory in $D_{i+1}$ dimensions.

Comparing with \eqref{vrhoi}, we see that this $v$-vector corresponds to a $w$-vector of
\be
w_{i}^{\text{Csmr}} = \begin{cases}
        0 & \text{if $j \leq  i$}  \\
           D_{j} &  \text{if $j > i$}\,.
		 \end{cases} 
         \label{wCsmr}
\ee
In the case of a stringy phase, we must also set $v_\phi = 0$, since the Casimir energy is independent of the $D$-dimensional dilaton. Using \eqref{vphi} and setting $w_D =D$ from \eqref{wCsmr}, this implies
\be
k_g = 0\,.
\ee
Thus, Casimir energies also fall into our taxonomy program, as they are described by an integral $w$-vector and an integral $k_g = 0$.

\subsection{Dot product rules and lattices}\label{ss.Dotproduct}

From \eqref{vrhoi} and \eqref{vphi}, we may obtain simple radion and dilaton lattice rules analogous to those of \eqref{radionpr} and \eqref{dilatonpr}. In our basis, the $\alpha$-vector associated with the KK-modes of the $i$th compactification manifold takes the form
\be
(\vec \alpha_{i})_j = \begin{cases}
			0 & \text{if $i < j$}\\
             \sqrt{\frac{D_{i+1}-2}{(D_{i+1}-D_i)(D_{i}-2)}} & \text{if $i=j$}  \\
            \sqrt{\frac{D_{j+1}-D_j}{(D_{j+1}-2)(D_{j}-2)}} & \text{if $j < i$}\,. 
		 \end{cases} 
\ee
The magnitude of this vector is given by
\be
|\vec \alpha_{i}| = \sqrt{\frac{d+n_i-2}{n_i(d-2)}}\,,
\ee
where $n_i=D_{i+1}-D_i$ is the dimensionality of the $i$th compactification manifold. Thus, the unit vector in the $\vec \alpha_i$ direction is given by
\be
\hat \alpha_i = \frac{\vec \alpha_i}{|\vec \alpha_i|}= \sqrt{\frac{n_i(d-2)}{n_i+d-2}} \vec \alpha_i\,.
\ee
Taking the dot product between this unit vector and a $v$-vector given by \eqref{vrhoi}, we find
\begin{align}
\vec v \cdot \hat \alpha_i =  \sqrt{\frac{n_i(d-2)}{n_i+d-2}} \Bigg[ &\sum_{j=1}^{i-1} \frac{(w_{j+1} - 2)(D_{j}-2) - (w_{j}-2)(D_{j+1}-2) }{\sqrt{(D_{j+1}-D_{j})(D_{j+1}-2)(D_{j}-2)}} \sqrt{\frac{D_{j+1}-D_j}{(D_{j+1}-2)(D_{j}-2)}} \nonumber \\
&+ \frac{(w_{i+1} - 2)(D_{i}-2) - (w_{i}-2)(D_{i+1}-2) }{\sqrt{(D_{i+1}-D_{i})(D_{i+1}-2)(D_{i}-2)}} \sqrt{\frac{D_{i+1}-2}{(D_{i+1}-D_i)(D_{i}-2)}}\Bigg] \,.
\end{align}
The sum over $j$ is a telescoping sum, and the whole expression collapses to
\begin{align}
\vec v \cdot \hat \alpha_i = \sqrt{\frac{n_i(d-2)}{n_i+d-2}} \left(  \frac{w_{i+1}-w_i}{D_{i+1}-D_i}  - \frac{w_1-2}{D_1-2}\right)\,.
\end{align}
By definition, we have $D_1=d$, $w_1 = 0$, and the dimension of the compactification manifold is related to the dimensionality of spacetime by $n_i = D_{i+1} -D_i$. Let us finally define the integer $k_i$ by
\be
k_i = n_i - (w_{i+1}-w_i)\,.
\label{kiint}
\ee
The meaning of this quantity will become clear shortly. With this and a bit of algebra, our expression simplifies to 
\begin{align}
\vec v \cdot \hat \alpha_i = \frac{d(n_i+d-2)-(d+k_i)(d-2)}{\sqrt{n_i(n_i+d-2)(d-2)}}\,.
\end{align}
Comparing with \eqref{radionpr}, we see that this precisely matches the radion lattice rule of a codimension-0 brane in $d$ dimensions. If the potential comes from a codimension-0 brane, the integer $k_i$ can be interpreted as the number of dimensions wrapped by the brane around the $i$th compactification manifold.

From our discussion in \S\ref{ss.weylweights}, we expect that in general, $w_{i+1} - w_i \geq -2$. From \eqref{kiint}, this yields an upper bound $k_i \leq n_i  +2$.

Next, we consider dilaton dot product rules. The $\alpha$-vector associated with a tower of string oscillator modes takes the form
\be
(\vec \alpha_{\rm osc})_i = \begin{cases}
        -\frac{1}{\sqrt{D-2}} & \text{if $i=N$}  \\
            \sqrt{\frac{D_{i+1}-D_i}{(D_{i+1}-2)(D_{i}-2)}} & \text{if $i < N$}\,. 
		 \end{cases} 
\ee
This vector has length $|\vec \alpha_{\rm osc}| = 1/\sqrt{d-2}$, so $\hat \alpha_{\rm osc} = \sqrt{d-2}\vec \alpha_{\rm osc} $. Taking the dot product with $\vec v$ and using \eqref{vrhoi}, \eqref{vphi}, we find
\begin{align}
\vec v \cdot \hat \alpha_{\rm osc} =  \sqrt{d-2} \Bigg[ \sum_{i=1}^{N-1} &\frac{(w_{i+1} - 2)(D_{i}-2) - (w_{i}-2)(D_{i+1}-2) }{\sqrt{(D_{i+1}-D_{i})(D_{i+1}-2)(D_{i}-2)}} \sqrt{\frac{D_{i+1}-D_i}{(D_{i+1}-2)(D_{i}-2)}} \nonumber \\
&- \frac{1}{D-2} \left( w_N - D - \frac{k_g(D-2)}{2} \right) \Bigg]\,,
\end{align}
where $w_D =w_N$, $D=D_N$. The sum over $i$ again yields a telescoping sum. Setting $w_1 = 0$, $D_1=d$, $D_N=D$, we find
\be
\vec v \cdot \hat \alpha_{\rm osc} = \frac{d}{\sqrt{d-2}} + \frac{\sqrt{d-2}}{2} k_g \,.
\ee
This yields a perfect match with the dilaton lattice rule \eqref{dilatonpr} for a codimension-0 brane, which has $p=d$, if we set \be
k_{\rm osc}= -k_g\,.
\label{poscvskg}
\ee
Since $k_g \geq -2$, we further have the restriction $k_{\rm osc} \leq 2$, whereas there is no lower bound on $k_{\rm osc}$.

The upshot of this is that the lattice of $v$-vectors precisely matches the lattice generated by $\alpha$-vectors of codimension-0 branes.

\subsubsection{$2\Gamma_{d-1} \subset \Gamma_d$}\label{sss.latticecontainment}

In \S\ref{sec:LEADING}, we saw that some potential terms scale with the tension of a codimension-0 brane, $V\sim T_{d}$, while others scale with the square of the tension of a codimension-1 brane, $V\sim T^2_{d-1}$. In \S\ref{ss.Dotproduct}, however, we argued that all $v$-vectors lie in the lattice $\Gamma_d$ generated by $\alpha$-vectors of codimension-0 branes. These claims are consistent only if the $\alpha$-vector of any codimension-1 brane satisfies $2\vec \alpha_{d-1} \in \Gamma_d$, or equivalently, if the lattice $\Gamma_{d-1}$ generated by these $\alpha$-vectors satisfies the containment relation $2\Gamma_{d-1} \subset \Gamma_d$. We now show that this is indeed the case.

Consider first the radion lattice condition for $\alpha$-vectors of codimension-0 and codimension-1 branes. By \eqref{radionpr}, these take the form:
\begin{subequations}
\begin{align}
    \hat \alpha_\text{KK}\cdot (2\vec \alpha_{d-1})&=2\frac {(d-1)(D-2)-P_{d-1}(d-2)}{\sqrt{(D-d)(D-2)(d-2)}},\label{e.cod1rad}\\
    \hat \alpha_\text{KK}\cdot \vec \alpha_d&=\frac{d(D-2)-P_d(d-2)}{\sqrt{(D-d)(D-2)(d-2)}}\,,
    \label{e.cod0rad}
\end{align}
\end{subequations}
for some integers $P_{d-1}$, $P_d$.

Setting
\begin{align}
    P_d\rightarrow2 - D + 2 P_{d-1}\,,
    \label{Pradsub}
\end{align}
we see that these expressions become equivalent. This immediately shows that in a Planckian phase, the vector $2 \vec \alpha_{d-1}$ lies in $\Gamma_d$.

In the case of a stringy phase, we must also consider the dilaton lattice rules. By \eqref{dilatonpr}, we have
\begin{subequations}
\begin{align}
\hat \alpha_\text{osc}\cdot (2\vec \alpha_{d-1})&=2\left(\frac{d-1}{\sqrt{d-2}}+\frac{\sqrt{d-2}}2(d-1-P_{d-1})\right)\,,\label{e.cod1dil}\\
    \hat \alpha_\text{osc}\cdot \vec \alpha_d&=\frac{d}{\sqrt{d-2}}+\frac{\sqrt{d-2}}2(d-P_d)\,.\label{e.cod0dil}
\end{align}
\end{subequations}
By setting
\begin{align}
    P_d\rightarrow 2 P_{d-1}-d\,,
    \label{Pphisub}
\end{align}
we again find that these expressions become equivalent.

Together, these results for both the radion and dilaton lattices imply that twice the codimension-1 lattice is a sublattice of the codimension-0 lattice,
\begin{align}
   2 \Gamma_{d-1} \subseteq \Gamma_d\,.
\end{align}
Note that this is, in general, a proper containment. In particular, if $P_d+d$ is odd, then the substitution \eqref{Pphisub} would require a half-integral value of $P_{d-1}$, indicating that this point does not lie in $2 \Gamma_{d-1}$. Similarly, \eqref{Pradsub} cannot be satisfied for integral $P_{d-1}$ if $D+P_d$ is odd.

\section{Relation to Swampland Bounds}\label{sec:Relation}

\subsection{Strong Asymptotic de Sitter Conjecture}\label{ss.STRONG}

In \cite{Rudelius:2021oaz, Rudelius:2021azq, Etheredge:2022opl, Rudelius:2022gbz}, compelling evidence was provided for the Strong Asymptotic de Sitter Conjecture, which holds that positive potentials in the asymptotic regime of scalar field space in quantum gravity satisfy
\be
\frac{|\nabla V|}{V} \geq \frac{2}{\sqrt{d-2}}\,.
\label{strongdSC}
\ee
This bound may be viewed as a sharpened version of the de Sitter Conjecture of \cite{Obied:2018sgi} with the important caveat that it applies only in asymptotic limits of scalar field space, not in the interior. The bound \eqref{strongdSC} is closely related to the strong energy condition, accelerated expansion \cite{Calderon-Infante:2022nxb, Andriot:2023wvg}, the existence of cosmological event horizons \cite{Hassfeld:2025hjx}, and (under certain assumptions) the TCC bound \cite{Bedroya:2019snp}.

In theories with multiple potential terms, the computation of $|\nabla V|/V$ can be nontrivial. 
Nonetheless, if all potential terms are non-negative (i.e., if $V_a \geq 0$ for all $a$), then a sufficient condition for \eqref{strongdSC} is the existence of a unit vector $\hat \lambda$ such that all terms $V_a$ satisfy
\be
(- \vec \nabla \log V_a) \cdot \hat \lambda \geq \frac{2}{\sqrt{d-2}}\,. \label{hatlambdacond}
\ee
To see this, we write the full potential as a sum of terms in our taxonomy program as
\be
V_{\rm tot} = \sum_{a} V_a\,.
\label{Vtot}
\ee
We then have
\be
\frac{\vec \nabla V_{\rm tot}}{V_{\rm tot}} = \sum_a \frac{\vec \nabla V_{a}}{V_{a}} \frac{V_a}{V_{\rm tot}} = \sum_a \vec v_a x_a\,,
\ee
where
\be
x_a = \frac{V_a}{V_{\rm tot}} \,,~~~~ \sum_a x_a = 1\,.
\ee
Assuming \eqref{hatlambdacond}, we then have
\be
|\vec \nabla \log V_{\rm tot}| \geq |\hat \lambda \cdot \frac{\vec \nabla V_{\rm tot}}{V_{\rm tot}}| = |\sum_a (\hat \lambda \cdot \vec v_a) x_a| \geq \frac{2}{\sqrt{d-2}} \sum_a x_a = \frac{2}{\sqrt{d-2}}\,,
\label{dSCmultiterm}
\ee
as desired. Note that in the second equality in \eqref{dSCmultiterm}, we used the assumption that all $V_a$ are non-negative, hence $x_a \geq 0$. In general, some potential terms $V_{\bar a}$ can be negative without affecting the conclusion provided that they are subleading, which means that $V_{\bar a}$ decays at a faster rate than the full potential $V_{\rm tot}$, so $x_{\bar a} \rightarrow 0$ in the asymptotic limit. 

In what follows, we will show that \eqref{hatlambdacond} is satisfied for any potential term $V_a$ satisfying the taxonomy rules when $\hat \lambda$ is chosen to point in the direction of the species vector, defined above in \eqref{speciesvector}:
\be
\hat \lambda = \frac{\vec{\lambda}}{|\vec \lambda|}\,,~~~~ \vec \lambda = - \vec \nabla \log \Lambda_{\rm QG} \,.
\ee
Here, $\Lambda_{\rm QG}$ represents the species scale, which can be parametrically identified with either the string scale or the Planck scale of the parent theory before compactification. We address each of these cases in turn.

\subsubsection{Stringy phase}\label{sss.dSCStringy}

In a stringy phase, the species scale may be identified with the fundamental string scale of the parent theory before dimensional reduction. Using the notation of \S\ref{sec:REVIEW}, the corresponding species vector takes the form \cite{Etheredge:2024tok}:
\be
\vec \lambda = \left( \sqrt{\frac{D_2-D_1}{(D_1-2)(D_2-2)}}, ...,\sqrt{\frac{D_{i+1}-D_{i}}{(D_i-2)(D_{i+1}-2)}} , ..., \sqrt{\frac{D_{N}-D_{N-1}}{(D_{N-1}-2)(D_{N}-2)}}, -\frac{1}{\sqrt{D_N-2}} \right) \,.
\label{lambdavec}
\ee
Recall that the final component of this vector measures the dilaton-dependence, while the first $N$ components measure the radion-dependence for the respective sequence of decompactifications.
The length of this vector is given simply by $|\vec \lambda| = 1/\sqrt{d-2}$, where $d=D_1$.

Now, let us consider the dot-product between this and the $v$-vector associated with a given potential term. In \eqref{sec:GENERAL}, we saw that these terms are characterized by a vector of integers $\vec w = (w_1, ..., w_N)$ and an integer $k_g$ that measures the scaling behavior of the term with the string coupling $g_s$.

The radion components of the associated $v$-vector are given by \eqref{vrhoi}:
\be
v_{i} = \sqrt{\frac{1}{(D_{i+1}-D_{i})(D_{i+1}-2)(D_{i}-2)}}\Big( (w_{i+1} - 2)(D_{i}-2) - (w_{i}-2)(D_{i+1}-2) \Big) 
\,,
\ee
where $i=1,...,N -1$.

The remaining dilaton component of the $v$-vector is given by \eqref{vphi}:
\be
v_N = v_\phi  =  \frac{1}{\sqrt{D_N-2}} \left( -\frac{k_g (D_N-2)}{2} - D_N +  w_N \right)    \,.
\label{vphi2}
\ee

Taking the dot product $\vec \lambda \cdot \vec v$, we find
\begin{align}
\vec \lambda \cdot \vec v &= \frac{-w_N + \frac{1}{2}(D_N-2) k_g + D_N}{D_N-2} + \displaystyle\sum_{i=1}^{N-1} \frac{(w_{i+1} - 2)(D_{i}-2) - (w_{i}-2)(D_{i+1}-2)}{(D_{i+1}-2)(D_{i}-2)} \nonumber \\ &=
 \frac{-w_N + \frac{1}{2}(D_N-2) k_g + D_N}{D_N-2} + \displaystyle\sum_{i=1}^{N-1} \left( \frac{w_{i+1} - 2}{D_{i+1}-2} - \frac{w_{i} - 2}{D_{i}-2} \right) \,.
\end{align}
Remarkably, this is a telescoping sum. Canceling out consecutive terms, we are left simply with
\begin{align}
\vec \lambda \cdot \vec v &=\frac{-w_N + \frac{1}{2}(D_N-2) k_g + D_N}{D_N-2}  + \frac{w_N - 2}{D_N-2} - \frac{w_1 - 2}{D_1-2} \nonumber  \\
&=  1+  \frac{k_g}{2} + \frac{2}{d-2}\,,
\end{align}
where in the last step we used the fact that $w_1 =0$ for a term in the potential and set $D_1 =d$.
Finally, we use the fact that $k_g \geq -2$ in string perturbation theory, since terms in the closed string tree-level sector scale as $g_s^{-2}$, and all other sectors come with larger powers of $g_s$.\footnote{Similarly, the tension of an NS5-brane \eqref{TNS5} scales as $T_{\rm NS5} \sim g_s^{-2}$ and thus has $k_g = -2$; all other branes in string theory have $k_g > -2$.} This implies $\vec \lambda \cdot \vec v \geq 2/(d -2)$, and since $|\vec \lambda|= 1/\sqrt{d-2}$, we finally conclude
\be
\vec{v} \cdot \hat \lambda = \frac{\vec \lambda \cdot \vec v}{|\vec \lambda|} \geq \frac{2}{\sqrt{d- 2}} \,,
\label{vdotlamstringy}
\ee
which is the desired result. Therefore, in a stringy phase, the Strong Asymptotic de Sitter Conjecture follows from our taxonomy rules for sums of positive potential terms together with the observation that string perturbation theory begins at order $g_s^{-2}$.

\subsubsection{Planckian phase}\label{sss.dSCPlanckian}

The case of a Planckian phase is very similar to the case of the stringy phase considered above. The key distinction is that there is no dilaton in the parent theory in $D=D_N$ dimensions, which effectively removes the final component of the $\lambda$-vector \eqref{lambdavec} and the final component of the $v$-vector \eqref{vphi2}. Consequently, the length of the $\lambda$-vector is reduced to \cite{Etheredge:2024tok}:
\be
|\vec \lambda| = \sqrt{\frac{D - d}{(D - 2) (d-2)}} \,.
\ee
The dot-product $\vec \lambda \cdot \vec v$ is identical to that of the stringy phase, excluding the dilatonic term:
\begin{align}
\vec \lambda \cdot \vec v &= + \displaystyle\sum_{i=1}^{N-1} \frac{(w_{i+1} - 2)(D_{i}-2) - (w_{i}-2)(D_{i+1}-2)}{(D_{i+1}-2)(D_{i}-2)} \nonumber \\ &=
\displaystyle\sum_{i=1}^{N-1} \left( \frac{w_{i+1} - 2}{D_{i+1}-2} - \frac{w_{i} - 2}{D_{i}-2} \right) = \frac{w_N - 2}{D-2} + \frac{2}{d-2} \,,
  \label{Planckianlv}
\end{align}
where in the last step we have condensed the telescoping sum, set $D_1=d$, $D_N=D$, and used the fact that $w_1 = 0$ for any term in the potential.

At this point, let us use our familiarity with M-theory to assume that the parent theory in $D$ dimensions does not have a cosmological constant. This ensures that there is no term with $w_N = 0$, so $w_N \geq 1$. With this assumption, we have
\be
\vec{v} \cdot \hat \lambda = \frac{\vec \lambda \cdot \vec v}{|\vec \lambda|} \geq \sqrt{\frac{(D - 2) (d-2)}{D-d} \left( \frac{-1}{D-2} + \frac{2}{d-2} \right)} \,.
\ee
Finally, a straightforward algebraic manipulation reveals that for $D > d > 2$, the right-hand side of this equation is bounded below by $2 /\sqrt{d - 2}$, with saturation occurring in the limits $D \rightarrow \infty$ or $d \rightarrow 2^+$. We therefore conclude that for $d > 2$,
\be
\vec{v} \cdot \hat \lambda \geq \frac{2}{\sqrt{d-2}}\,,
\label{vdotlamPlanckian}
\ee
as desired.

Note that this equation will be violated if we set $w_N = 0$ in \eqref{Planckianlv}, corresponding to a cosmological constant in the parent theory in $D$ dimensions. This is a sensible result: the Strong Asymptotic de Sitter Conjecture was motivated by the idea that, at late times, the causal structure of a pocket universe should approach that of an asymptotically decelerating universe rather than a de Sitter spacetime \cite{Rudelius:2021azq}. A positive cosmological constant in the parent $D$-dimensional theory would imply something even more surprising: a stable de Sitter vacuum. Longstanding quantum gravity lore forbids such vacua \cite{Goheer:2002vf}.

To conclude this subsection, we stress once again that the bound $|\vec \nabla \log V| \geq 2/\sqrt{d-2}$ we have derived here applies only in asymptotic regimes of scalar field moduli space, where we generically expect our taxonomy rules to apply. Debates regarding the existence and construction of de Sitter minima in the interior of scalar field moduli space in quantum gravity are ongoing (see e.g. \cite{Ooguri:2018wrx, Danielsson:2018ztv,Andriot:2019wrs, McAllister:2024lnt}), and we do not address them here.

\subsection{Decay rates of potentials vs. mass scales \label{ss.massvspotential}}

A scalar field rolling in an asymptotic potential with $V > 0$ generates an expanding FRW cosmology at late times. As the field rolls, the potential $V$, the Hubble scale $H$, the species scale $\Lambda_{\rm QG}$, and the masses of the principal towers $m_i$ all decay exponentially with geodesic distance $||\phi||$:
\be
V \sim \exp(-c ||\phi||)\,,~~~~H \sim \exp(-\lambda_H ||\phi||)\,,~~~~\Lambda_{\rm QG} \sim \exp(-\lambda ||\phi||)\,,~~~~m_i \sim \exp(- \alpha_i ||\phi||)\,.
\ee
The decay rates can be computed from the relevant $v$-vectors, $\alpha$-vectors, and $\lambda$-vector. Consider a generic infinite-distance limit $\vec t=\vec \phi_0+\phi \hat t$ in the $\hat t$ direction of the principal plane. In such a limit, a principal tower of mass $m_i$, the species scale $\Lambda_{\rm QG}$, and a potential term $V$ will scale as
\begin{align}
m_i(\phi)\sim \exp\left(-\vec \alpha_i \cdot \hat t\ \phi\right),\qquad \Lambda_{\rm QG} \sim \exp(- \vec \lambda \cdot \hat t\ \phi), \qquad V(\phi)\sim \exp\left(-\vec v \cdot \hat t\ \phi\right),
\end{align}
where $\vec \alpha_i$ is the $\alpha$-vector of the principal tower, $\vec \lambda$ is the species vector, and $\vec v$ is the $v$-vector of the potential term. 
Hence, we have 
\be
c = \vec v \cdot \hat t,~~~~ \lambda = \vec \lambda \cdot \hat t, ~~~~\alpha_i = \vec \alpha_i \cdot \hat t\,.
\label{clambdalpha}
\ee
If there are multiple potential terms $V_a$, the potential decay rate $c$ is given by the minimal value of $\vec v_a \cdot \hat t$ over all $a$.

Along a rolling scalar field trajectory, the decay rate of the Hubble scale $\lambda_H$ is given in terms of the decay rate of the potential $c$ by \cite{Rudelius:2022gbz}:
\be
\lambda_H = \min\left(\frac{c}{2} ,\sqrt{\frac{d-1}{d-2}} \right)\,. 
\label{lambdaH}
\ee
Furthermore, if $\frac{c}{2} < \sqrt{\frac{d-1}{d-2}}$, then at late times the scalar field trajectory will approach a path of gradient descent. If $\frac{c}{2} > \sqrt{\frac{d-1}{d-2}}$, on the other hand, then at late times the scalar field will undergo kination \cite{Joyce:1996cp, Gouttenoire:2021jhk}, and the potential will play a negligible role in its dynamics.

In this subsection, we will use \eqref{clambdalpha} and \eqref{lambdaH} together with our taxonomy rules to compare the decay rates of various mass scales.

\subsubsection{Hubble vs. species scale}\label{Hubvsspecies}

As argued in e.g. \cite{Hebecker:2018vxz}, consistency of effective field theory imposes a restriction on the Hubble scale and the species scale:
\be
H < \Lambda_{\rm QG} \,.
\ee
This follows from the fact that the species scale serves as a UV cutoff on the EFT, while the Hubble scale serves as an IR cutoff. This in turn requires
\be
\lambda_H \geq \lambda \,,
\ee
so that the Hubble scale decays more quickly than the species scale.

For a potential of the form $V_{\rm tot}= \sum_a V_a$ where all leading terms have $V_a \geq 0$, this relation follows from the result of \S\ref{ss.STRONG}.
 To see this, we first note that \eqref{clambdalpha} and \eqref{lambdaH} yield
\be
\lambda_H = \min_a \left(\frac{1}{2} \vec v_a \cdot \hat t, \sqrt{\frac{d-1}{d-2}} \right)\,,
\ee
where $\hat t$ is the direction of gradient descent in the asymptotic potential and $a$ runs over the terms in the asymptotic potential. From 
\eqref{vdotlamstringy} and \eqref{vdotlamPlanckian}, we then have that for all $a$,
\be
\frac{1}{2} \vec v_a \cdot \hat \lambda \geq \frac{1}{\sqrt{d-2}}\,.
\ee
If $\hat t$ is taken to be the direction of gradient descent (i.e., the direction where $(- \vec \nabla \log V) \cdot \hat t$ is maximized), we further have 
\be
\min_a (\vec v_a  \cdot \hat t) \geq \min_a (\vec v_a  \cdot \hat \lambda) \geq \frac{2}{\sqrt{d-2}} \,.
\ee
Finally, recalling that
$|\vec \lambda| = 1/\sqrt{d-2}$ in a stringy phase, while $|\vec \lambda| < 1/\sqrt{d-2}$ in a Planckian phase \cite{vandeHeisteeg:2023ubh,Calderon-Infante:2023ler, Etheredge:2024tok}, we conclude that
\be
\lambda_H = \min_a \left(\frac{1}{2} \vec v_a \cdot \hat t, \sqrt{\frac{d-1}{d-2}} \right) \geq \frac{1}{\sqrt{d-2}}  \geq |\vec \lambda| \geq \lambda \,.
\ee
This establishes the desired result $\lambda_H \geq \lambda$, showing that the Hubble scale decays at least as quickly as the species scale along such a trajectory.

\subsubsection{Hubble vs. tower masses}\label{sss.Hubvstower}

As a scalar field rolls in an asymptotic regime, the underlying theory will generically decompactify and approach weak string coupling, as both $m_{\rm KK}$ and $m_{\rm osc}$ decay exponentially relative to the Planck scale. In order to maintain a $d$-dimensional FRW cosmology during this process, the Hubble scale must satisfy
\be
H \lesssim m_{\rm KK}\,, m_{\rm osc}\,.
\label{Hbound}
\ee
In particular, if $H > m_{\rm KK}$, then the horizon is smaller than the size of the extra dimensions, and the theory cannot be properly called a $d$-dimensional cosmology. If $H > m_{\rm osc}$, then effective field theory breaks down, as the species scale (a UV cutoff) lies below the Hubble scale (an IR cutoff) \cite{Hebecker:2018vxz}. 

Since $m_{\rm osc} \sim \Lambda_{\rm QG}$ in a stringy phase, the latter requirement $H \lesssim m_{\rm osc}$ follows from the result of \S\ref{Hubvsspecies}. By \eqref{clambdalpha}, the former requirement $H \lesssim m_{\rm KK}$ implies that along an asymptotic trajectory in the $\hat t$ direction, we must have 
\be
\vec v_a \cdot \hat t \geq 2 \vec \alpha_i \cdot \hat t ~~~\text{for all $a$, $i$} \,,
\label{vavsalphai}
\ee
where $a$ runs over the terms of the potential (with associated $v$-vectors $\vec v_a$) and $i$ runs over the principal towers (with associated $\alpha$-vectors $\vec \alpha_i$).

In theories with multiple competing potential terms, the computation of $\hat t$ can be complicated. However, in theories with just a single term in the potential, the gradient descent direction $\hat t_{\rm GD}$ is given by the direction of the $v$-vector of that potential term:\footnote{For potentials that decay sufficiently quickly ($c > 2 \sqrt{(d-1)/(d-2)}$), the asymptotic direction of motion of the scalar field may differ from the direction of gradient descent $\hat t_{\rm GD}$ depending on its initial conditions, as the gradient descent trajectory is not a late-time attractor \cite{Copeland:1997et, Tsujikawa:2013fta, Rudelius:2022gbz}. However, there always exist trajectories whose asymptotic direction is given by $\hat t_{\rm GD}$. Consequently, \eqref{v2vs2valpha} may be viewed as a necessary condition in this case.}
\be
\hat t_{\rm GD} = \frac{\vec v}{|\vec v|}\,.
\ee
With this, \eqref{vavsalphai} becomes
\be
\vec v \cdot \vec v \geq 2 \vec v \cdot \vec \alpha_i ~~~ \text{for all $i$}\,.
\label{v2vs2valpha}
\ee
More generally, in theories with multiple potential terms $V_a$, \eqref{vavsalphai} implies that \eqref{v2vs2valpha} is a necessary condition for any term in the asymptotic potential.
In what follows, we briefly explore this condition in the context of our taxonomy program, leaving a more detailed analysis to future work.

For simplicity, we focus our attention on the Kaluza-Klein tower associated with the first compactification manifold by setting
\be
\vec \alpha_i = \vec \alpha_1 = \left(\sqrt{\frac{d+n_1-2}{n_1(d-2)}}, 0, ... , 0\right)\,.
\ee
In the absence of sliding \cite{Etheredge:2023odp} (see \S\ref{sec:REVIEW}), this may be done without loss of generality, as we can rearrange the order of decompactifications by varying the direction of the infinite-distance limit of interest. More generally, the requirement $\vec v \cdot \vec v \geq 2 \vec v \cdot \vec \alpha_1$ may be viewed as a necessary but not necessarily sufficient condition to establish \eqref{v2vs2valpha}.

From \eqref{vrhoi}, we have
\be
v_1 = \frac{(w_2-2)(d-2)+2(n_1+d-2)}{\sqrt{n_1(n_1+d-2)(d-2)}}\,,
\ee
where as usual we have set $D_1=d$, $D_2=n_1+d$, and $w_1 =0$.

We then have
\begin{align}
 \vec v^2 - 2 \vec v\cdot \vec \alpha_1 & \geq v_1^2 - 2 \vec v\cdot \vec \alpha_1  \nonumber \\
 &= (w_2-2) \frac{w_2 (d-2) + 2 n_1}{n_1(n_1+d-2)}\,.
 \label{v2inequal}
\end{align}
Since $d >2$, $w_2 \geq 0$, and $n_1 > 0 $, we conclude that \eqref{v2vs2valpha} is satisfied for $i=1$ provided
\be
w_2 \geq 2\,.
\ee
From \eqref{sec:GENERAL}, the only potential terms with $w_2 = 0$ are those that come from potential terms in $d+n_1$ dimensions, and the only terms with $w_2=1$ are those that come from codimension-1 branes in $d+n_1$ dimensions that are wrapped over $n_1-1$ dimensions to produce codimension-0 branes in $d$ dimensions. For all other sources of potential terms, we conclude that $\vec v^2 \geq 2 \vec v \cdot \vec \alpha_1$.

But what about potentials descending from higher-dimensional potentials or codimension-1 branes? Do these present counterexamples to \eqref{v2vs2valpha}? If the moduli space dimension is greater than 1, then the inequality in the first line of \eqref{v2inequal} will generically be a strict inequality. Further analysis is needed to determine whether $w_2 < 2$ leads to a violation of \eqref{v2vs2valpha}.

If the moduli space is 1-dimensional, however, then the inequality in the first line of \eqref{v2inequal} is saturated, and an instance of $w_2 < 2$ would present a counterexample to \eqref{v2vs2valpha}. This scenario requires a Planckian phase with a single compactification manifold, of which a compactification of M-theory is the only familiar example. As discussed in \S\ref{ss.STRONG}, M-theory does not have a cosmological constant, which excludes the possibility of $w_2 =0$. However, M-theory \emph{does} have a codimension-1 brane: namely, the Hořava-Witten wall (also known as an M9-brane). If one could obtain a potential in 10d from an unwrapped M9-brane, this potential would violate \eqref{v2vs2valpha}.

However, crucially, the M9-brane is not a simple dynamical object in the way that the M2-brane and M5-brane are. Rather, it appears as a boundary condition on the M-theory interval, and the resulting heterotic string theory in 10d does not receive a vacuum energy from the presence of the M9-brane. Thus, this candidate counterexample to \eqref{v2vs2valpha} is not realized within the known landscape. We leave a more thorough investigation to future work.

\subsubsection{Potential vs. lightest tower}\label{sss.potvslight}

To conclude this subsection, we derive a sort of converse to the bound in \S\ref{sss.Hubvstower}. In the process, we record a number of identities relating $v$-vectors and $\alpha$-vectors of principal towers.

In \S\ref{sec:GENERAL}, we saw that the lattice of $v$-vectors can be identified with the lattice of $\alpha$-vectors of codimension-0 branes.
Using the taxonomy results of \cite{Etheredge:2025ahf} for brane $\alpha$-vectors, we may then write any $v$-vector as a linear sum of $\alpha$-vectors of principal towers,
\begin{align}
\vec v=\sum_i\gamma_i\ \vec \alpha_i\,.
\end{align}
The coefficients $\gamma_i$ are fixed by the $v$-vector residing on the lattice of $\alpha$-vectors for codimension-0 branes. Explicitly, by (1.9) of \cite{Etheredge:2025ahf}, the $\gamma_i$ are given in a Planckian phase by
\begin{align}
\gamma_i=\frac{Pn_i}{D-2}-k_i\,,
\end{align}
where here and below we have adopted the notation of \S\ref{sec:REVIEW} for the taxonomy of codimension-0 branes.

Similarly, by (1.12) of \cite{Etheredge:2025ahf}, the $\gamma_i$ are given in a stringy phase by
\begin{align}
\gamma_\text{KK,$i$}=\frac{n_ik_\text{osc}}2-k_i,\qquad \gamma_\text{osc}=\tilde P-\frac{k_\text{osc}(\tilde D-2)}2.
\end{align}

With these formulas, the potential term scales with the principal tower masses as
\begin{align}
V\sim \prod_im_i^{\gamma_i}.
\end{align}

Consider now a scalar field trajectory in the asymptotic regime of moduli space, and suppose further that the asymptotic potential is dominated by a single term $V$ with $v$-vector $\vec v$.
Consistency of the EFT description then implies that this $v$-vector must lie within the positive cone of the set of principal $\alpha$-vectors, i.e.,\footnote{In the language of \cite{Etheredge:2024tok}, this means that the $v$-vector lies in the \emph{\textbf{frame simplex}} of the given phase. If \eqref{gammai} were not satisfied, then far out along the geodesic there would be other, lighter principal towers, indicating that the original choice of duality frame was incorrect.}
\begin{align}
\gamma_i\geq 0.
\label{gammai}
\end{align}
Let $m_\text{lightest}$ be the lightest tower as one travels in the direction of $\vec v$. Then, as one travels in the direction of $\vec v$,
\begin{align}
V\gtrsim \prod_i m_\text{lightest}^{\gamma_i}=m_\text{lightest}^{\sum_i \gamma_i}.
\end{align}

For a Planckian phase,
\begin{align}
\sum_i \gamma_i=\sum_i\left(\frac{Pn_i}{D-2}-k_i\right)=d-\frac{P(d-2)}{D-2},
\end{align}
and so
\begin{align}
V\gtrsim m_\text{lightest}^{d-\frac{P(d-2)}{D-2}}  =  m_\text{lightest}^{\frac{w_N(d-2)+2(D-d)}{D-2}},\qquad\text{(Planckian),}\label{e.VmlightestPlanck}
\end{align}
where in the final step we have used the relation $P=D-w_N$ to express the bound using the notation of \S\ref{sec:GENERAL}.

For a stringy phase,
\begin{align}
\sum_i \gamma_i&=\sum_{i\neq \text{osc}}\left(\frac{k_\text{osc}n_i}2-k_i\right)+\tilde P-\frac{(\tilde D-2)k_\text{osc}}2=d-\frac{k_\text{osc}(d-2)}2.
\end{align}
Thus,
\begin{align}
V\gtrsim m_\text{lightest}^{d-\frac{k_\text{osc}(d-2)}2} = m_\text{lightest}^{d+\frac{k_g (d-2)}2},\qquad (\text{stringy}),\label{e.Vmlighteststringy}
\end{align}
where in the final step we have used the relation $k_{\rm osc}= -k_g$ to express the bound in the notation of \S\ref{sec:GENERAL}. Remarkably, this expression depends only on $k_g$ and not on the $w$-vector of Weyl weights $\vec w$.

An important caveat here is that the parameters $w_N$, $k_g$ appearing respectively in \eqref{e.VmlightestPlanck} and \eqref{e.Vmlighteststringy} are both bounded below, but they can be arbitrarily large. More precisely, terms with very large $w_N$ correspond to very high-derivative operators in the $D$-dimensional action, while terms with very large $k_g$ correspond to terms at high loop order in the $g_s$ expansion. Thus, while these bounds do not immediately yield universal lower bounds on the size of the asymptotic potential, they do yield meaningful bounds for a given order in a string-loop and Wilsonian expansion. If one could place an upper bound on $w_N$ and $k_g$ for the leading term in the asymptotic potential, our results here would immediately lead to a universal lower bound on $V$ in terms of the lightest tower scale $m_{\rm lightest}$. We leave this for future work.

\subsection{Species scale and Wilson coefficients}\label{ss.species}

We have seen that the asymptotic moduli-dependence of any term in the potential can be encoded by a $v$-vector, which is uniquely determined by a vector of integers $\vec{w}$ and (in a stringy phase) an integer $k_g$ determined by its order in the string loop expansion. Any term in the potential necessarily has $w_1 = 0$, ensuring that it transforms trivially under a Weyl transformation.

This formalism extends naturally from terms in the potential to more general terms in the $d$-dimensional effective action. Given such a term $\mathcal{O}$, we may once again define
\be
\vec{v}(\mathcal{O}) = - \vec \nabla \log \mathcal{O}\,,
\ee
where the gradient is taken with respect to the dilaton and the radions. This $v$-vector is again determined uniquely by the vector of Weyl weights $\vec{w}$ and the $g_s$-scaling coefficient $k_g$ through the formulas \eqref{vrhoi}, \eqref{vphi}. For a more general term in the action, we may however drop the requirement that $w_1 = 0$: such terms may transform nontrivially under Weyl transformations.

Let us focus our attention on higher-derivative gravitational corrections, which take the schematic form
\begin{equation}
    S = \frac{M_{\text{Pl};d}^{d-2}}{2} \int d^dx \sqrt{-g} \left( \mathcal{R}_d  
 + c_2 \mathcal{R}_d ^2 + c_3 \mathcal{R}_d^3 + ...\right)\,.
 \label{speciescoef}
\end{equation}
Here, the $c_i$'s are (moduli-dependent) Wilson coefficients. In \cite{vandeHeisteeg:2022btw, vandeHeisteeg:2023dlw, Castellano:2023aum}, it was argued that these coefficients should be suppressed by powers of the species scale $\Lambda_{\rm QG}$:\footnote{As discussed in e.g. \cite{Reece:2025zva}, this relation must rely on supersymmetry or a similar phenomenon, since in our universe, these coefficients are suppressed by the neutrino mass scale, which is many orders of magnitude below the species scale.}
\be
c_k = \frac{\tilde c_k}{\Lambda_{\rm QG}^{2k - 2}}\,, 
\label{cLQG}
\ee
where the $\tilde c_k$'s are order-one numbers. In what follows, we will present compelling evidence for this relation by computing the radion- and dilaton-dependence of each side.

We begin with the right-hand side of \eqref{cLQG}. We immediately have
\be
- \vec \nabla \log \left( \frac{\tilde c_k}{\Lambda_{\rm QG}^{2k - 2}} \right) = (2- 2k ) \vec{\lambda}_{\rm QG}\,,
\ee
where $\vec \lambda_{\rm QG} = - \vec \nabla \log \Lambda_{\rm QG}$ is the species vector, and we have assumed that any moduli-dependence of the order-one coefficient $\tilde c_k$ is negligible in the asymptotic limit. This species vector is given by \eqref{lambdavec}, so we have
\be
- \partial_{\rho_i} \log \left( \frac{\tilde c_k}{\Lambda_{\rm QG}^{2k - 2}} \right) = (2-2k) \sqrt{\frac{D_{i+1}-D_i}{(D_i-2)(D_{i+1}-2)}} \,,
\label{partialrhoctilde}
\ee
and in the case of a stringy phase,
\be
- \partial_\phi \log \left( \frac{\tilde c_k}{\Lambda_{\rm QG}^{2k - 2}} \right) = \frac{2k-2}{\sqrt{D-2}} \,.
\label{partialphictilde}
\ee

Meanwhile, a term of the form $\mathcal{R}^k$ has Weyl weight $w=2 k$ in any number of dimensions. Thus, such a term is described by a $w$-vector with
\be
w_i =  2 k\,,~~~~ i = 1, ..., N\,.
\ee
From \eqref{vrhoi}, this yields an associated $v$-vector of the form
\begin{align}
v_i \equiv - \partial_{\rho_i} c_k &= \sqrt{\frac{1}{(D_{i+1}-D_{i})(D_{i+1}-2)(D_{i}-2)}}\Big( (2 k - 2)(D_{i}-2) - (2k -2)(D_{i+1}-2) \Big) \nonumber \\
&= \sqrt{\frac{D_{i+1}-D_{i}}{(D_{i+1}-2)(D_{i}-2)}} \left(2 - 2 k  \right) \,,
\end{align}
which perfectly matches \eqref{partialrhoctilde}. 

In the case of a string phase, we must also consider the dilaton-dependence of $c_k$. By \eqref{vphi}, we have
\begin{align}
v_N \equiv - \partial_{\phi} c_k &= \frac{1}{\sqrt{D-2}} \left( -\frac{k_g (D-2)}{2} - D +  2 k \right) \,.
\end{align}
Assuming that the term $\mathcal{R}_D^{k}$ is generated at tree-level in the closed string expansion, we have $k_g = -2$, hence
\be
- \partial_{\phi} c_k = \frac{2k-2}{\sqrt{D-2}}\,,
\ee
which perfectly matches \eqref{partialphictilde}.

We conclude that whenever the rules of our taxonomy program hold and (in the case of a stringy phase) higher-derivative gravitational corrections are generated at tree-level in closed string theory:
\be
- \vec \nabla \log c_k =- \vec \nabla \log \left( \frac{\tilde c_k}{\Lambda_{\rm QG}^{2k - 2}} \right)\,,
\ee
where the gradient is taken with respect to the radions and dilaton. This yields very strong evidence for the proposed relation \eqref{cLQG}.

\section{Conclusions}\label{sec:CONC}

In this paper, we have argued that potentials asymptotically can be written as a sum of terms,
\begin{align}
    V=\sum_a V_a
\end{align}
where each term $V_a$ scales with the moduli as $V_a \sim \exp(- \vec v_a \cdot \vec \phi)$, and the vectors $\vec v_a$ reside in a lattice. 
Said differently, we have argued that the moduli-gradients of logarithms of potential terms, $- \vec \nabla \log V_a$, are lattice-valued.

In particular, we have argued that for a radion $\rho$ corresponding to decompactification from $d$ to $D$-dimensions, a contribution to the potential must scale as
\begin{align}
    V_a \sim \exp\left(-\frac{d(D-2)-P_\rho (d-2)}{\sqrt{(D-d)(D-2)(d-2)}}\rho\right)
\end{align}
for some integer $P_\rho$. Similarly, for a $d$-dimensional dilaton $\phi$, the potential scales as
\begin{align}
    V_a \sim \exp\left(\left(\frac{d}{\sqrt{d-2}}+\frac{\sqrt{d-2}}{2}(d-P_\phi)\right)\phi\right)
\end{align}
for some integer $P_\phi$. We have placed bounds on the integers $P_\rho$ and $P_\phi$ and have shown that under reasonable assumptions, these bounds imply (a) the Strong Asymptotic de Sitter Conjecture and (b) that higher-derivative gravitational corrections are suppressed by powers of the species scale. All of the examples in the landscape that we have checked in this paper satisfy our claims.

Our work leaves open a number of directions for future research. To begin, while we have placed some bounds on the lattice sites occupied by $v$-vectors, there are likely to be additional restrictions. It would be interesting to identify precisely which lattice sites are/are not occupied by $v$-vectors within the landscape.

Our results also imply relations between the exponential decay rates of potentials and mass scales, as the relative positions of the $v$-vectors and $\alpha$-vectors are both rigidly fixed to lie on lattices. We have argued that the Hubble scale in an FRW cosmology decays at least as quickly as the species scale.
We have further shown that if the Hubble scale remains lighter than the KK-scale and string scale, then the $\alpha$-vector of the lightest KK-mode/string oscillator tower must satisfy $\vec v^2\geq 2\vec \alpha_\text{lightest}\cdot \vec v$. However, we have found a potential counterexample to this relation for potentials generated by codimension-0 branes descending from codimension-1 branes in a higher-dimensional theory. It would be worthwhile to explore this potential counterexample in more depth to see if it can be realized in the landscape.

Our work also implies that potentials are asymptotically eigenfunctions of the moduli-space Laplacian. As was explained in \cite{Aoufia:2026ztl}, functions in moduli space that are exponential functions of the coordinates on the principal plane are eigenfunctions of the moduli space Laplacian. In particular, the eigenvalue of the Laplacian of the potentials in this paper are given by
\begin{align}
    \nabla^2 V=kV,\qquad k=v^2+\nabla^2 \log V,
\end{align}
where $\nabla^2 \log V$ involves the dot product between the $v$-vector of the potential and a site on the instantonic $\alpha$-vector lattice \cite{Aoufia:2026ztl}. It would be interesting to explore possible consequences of this result.

This work can be combined with moduli space quantum mechanics of \cite{Anchordoqui:2025izb, Anchordoqui:2026nit, Anchordoqui:2026dS}, using the taxonomy rules for moduli-space Laplacians of \cite{Aoufia:2026ztl}. In \cite{Anchordoqui:2026nit}, quantum mechanical particles propagating in moduli space with a potential were explored, where the relationships between the potential and moduli-space geometry controlled the behavior of the states. In such asymptotic limits, the rate of moduli-space volume decay and rate at which the decay-rate of a potential combine to control the behavior of the states. Given our taxonomy rules governing the rates at which potentials decay, together with rules governing the rates at which moduli-space volumes decay in \cite{Aoufia:2026ztl}, one could envision a taxonomy of quantum mechanical wavefunctions in asymptotic limits of moduli spaces.

One of the reasons the Sharpened Distance Conjecture \cite{Etheredge:2022opl} does not automatically follow from the Emergent String Conjecture \cite{Lee:2019wij} is because of the possibility of strong warping \cite{Etheredge:2023odp}. Such warping affects the decay rates of KK-mode masses and could, in principle, lead to a violation of the Sharpened Distance Conjecture. The effects of such warping were further investigated in \cite{Raucci:2026fzp}, where it was found that if the asymptotic potential of the higher-dimensional theory satisfies the Strong dS Conjecture bound \cite{Rudelius:2021azq, Rudelius:2022gbz}, then the Sharpened Distance Conjecture is satisfied (and saturated) by the KK-modes in the lower-dimensional theory. We have seen here that our taxonomy rules imply the Strong dS Conjecture, suggesting that the Sharpened Distance Conjecture is satisfied in further warped compactifications. This may provide an avenue for a classification of $\alpha$-vector sliding in the context of warped compactifications, though further work is needed to see how our potential taxonomy rules are themselves affected by strong warping.

\section*{Acknowledgements}

We are grateful for conversations with Ben Heidenreich, Alvaro Herraez, Joaquin Masias, Muthusamy Rajaguru, Matthew Reece, and Irene Valenzuela. The work of TR was supported in part by STFC through grant
ST/T000708/1 and by the Royal Society through grant RGS/R2/252603. The work of DL is supported by the German-Israel-Project (DIP) on Holography and the Swampland.

\appendix

\bibliography{ref}
\bibliographystyle{utphys}

\end{document}